\documentclass[journal]{IEEEtran}

\newcommand{\acceptednotice}{This work has been accepted for publication in IEEE Open J. Commun. Soc. DOI: 10.1109/OJCOMS.2026.3695929}

\usepackage[utf8]{inputenc}
\usepackage[T1]{fontenc}
\usepackage{cite}
\usepackage{amsmath,amssymb,amsfonts}
\usepackage{bm}
\usepackage{graphicx}
\usepackage{textcomp}
\usepackage{newunicodechar}
\newunicodechar{−}{\textminus}
\DeclareGraphicsExtensions{.png,.eps,.ps,.pdf}

\usepackage{booktabs}
\usepackage{array}
\usepackage{multirow}
\usepackage{makecell}
\usepackage{siunitx}
\usepackage{adjustbox}
\usepackage{float}
\usepackage{dblfloatfix}
\usepackage{subcaption}
\usepackage[normalem]{ulem}
\usepackage[dvipsnames,table]{xcolor}
\usepackage[ruled,vlined]{algorithm2e}
\usepackage{url}

\usepackage{hyperref}
\usepackage[noabbrev]{cleveref}
\crefname{figure}{Fig.}{Figs.}
\crefname{equation}{Eq.}{Eqs.}

\usepackage{fancyhdr}
\def\BibTeX{{\rm B\kern-.05em{\sc i\kern-.025em b}\kern-.08em
    T\kern-.1667em\lower.7ex\hbox{E}\kern-.125emX}}

\newcommand{\ULlatPad}[5]{%
\kern#1\relax%
\begin{tabular}[c]{@{}r@{\;\textbar\;}r@{\;\textbar\;}r@{\;\textbar\;}r@{}}%
#2 & #3 & #4 & #5%
\end{tabular}%
}

\newcommand{\slot}[2]{\makebox[#1][r]{#2}}

\newcommand{\ULlatPadWW}[6]{
\kern#1\relax%
\begin{tabular}[c]{@{} \slot{#2}{#3} @{\;\textbar\;} \slot{#2}{#4} @{\;\textbar\;} \slot{#2}{#5} @{\;\textbar\;} \slot{#2}{#6} @{}}
\end{tabular}%
}

\begin{document}

\title{Experimental Evaluation of Multi-Connectivity Strategies for Reliable Low-Latency Uplink Communications in Rural 5G Networks}

\author{Carlos S. Alvarez-Merino, Alejandro Ramirez-Arroyo, Rasmus Suhr Mogensen,
Morten V. Pedersen, Miguel Villanueva-Fernández, Emil J. Khatib, Sergio Fortes,
Raquel Barco, and Preben E. Mogensen%
\thanks{Corresponding author: Carlos S. Alvarez-Merino (e-mail: csam@uma.es).}%
\thanks{Carlos S. Alvarez-Merino, Emil J. Khatib, Sergio Fortes, and Raquel Barco are with the Telecommunication Research Institute (TELMA), Universidad de Málaga, E.T.S. Ingeniería de Telecomunicación, Bulevar Louis Pasteur 35, 29010 Málaga, Spain (e-mail: \{csam, emil, sfr, rbarco\}@uma.es).}%
\thanks{Alejandro Ramirez-Arroyo, Miguel Villanueva-Fernández, and Preben E. Mogensen are with the Department of Electronic Systems, Aalborg University, 9220 Aalborg, Denmark (e-mail: \{araar, mvf, pm\}@es.aau.dk).}%
\thanks{Rasmus Suhr Mogensen and Morten V. Pedersen are with NanoPing, 9220 Aalborg, Denmark (e-mail: \{rasmus, morten\}@nanoping.com).}%
\thanks{This work was partially funded by PAR 4/2024 University of Malaga and NanoPing ApS.}}

\maketitle
\thispagestyle{fancy}

\begin{abstract}
Reliable low-latency communication is a key requirement for mission-critical and mobile autonomous systems, including teleoperation, autonomous navigation, and real-time uplink-dominant telemetry applications. While commercial 5G networks often provide adequate downlink performance, uplink performance in rural deployments may be constrained by radio-resource limitations and uplink power-control mechanisms. This paper presents a comprehensive experimental evaluation of multi-connectivity strategies over commercial 5G Non-Standalone networks, based on measurement campaigns conducted in urban, suburban, and rural environments. The study analyzes per-packet uplink and downlink latency, packet loss, and radio-layer KPIs across two mobile network operators. The measurements indicate that latency and reliability cannot be inferred solely from coverage indicators such as RSRP. In coverage-constrained scenarios, performance appears to be strongly influenced by uplink power-limited operation and partially correlated impairments across operators. Several multi-connectivity strategies are evaluated, including link aggregation, switching-based policies, and conditional packet duplication. A Primary-Anchored Adaptive Failover (PAAF) framework is introduced to selectively activate redundancy based on radio, latency and service cost considerations. The results suggest that Partial Duplication (PD) approaches can approach the latency and reliability performance of full duplication while substantially reducing duplication overhead in the evaluated rural scenario. The proposed PAAF mechanisms are evaluated through offline replay of synchronized dual-operator traces, allowing switching, aggregation, full duplication, and PD to be compared under matched radio conditions within the evaluated commercial 5G NSA rural deployment.
\end{abstract}

\begin{IEEEkeywords}
5G, Multi-Connectivity, Low-Latency Communications, Reliability, Packet Duplication, Link Aggregation, Uplink Power Control
\end{IEEEkeywords}

\section{Introduction}

\IEEEPARstart{C}{ritical} communications underpin a wide range of applications such as industrial automation, remote monitoring, teleoperation, public safety, and emergency response, where strict requirements on latency, reliability, and service continuity must be met \cite{3GPP_TS_22_261_R17}. In these contexts, transient communication impairments, rather than sustained outages, often dominate system failures, as short-lived latency spikes or packet loss can destabilize control loops and compromise operational safety. 

Although 5G systems have introduced substantial improvements in spectral efficiency, peak data rates, and latency, particularly in enhanced Mobile Broadband (eMBB) scenarios, the performance gains have been predominantly realized in the downlink (DL) direction. Network densification, massive MIMO deployments, wider carrier bandwidths, and aggressive DL scheduling have collectively enabled multi-gigabit-per-second throughput under favorable propagation conditions \cite{3GPP_TS_38_300_R17}. However, the uplink (UL) remains comparatively constrained, both from a radio-resource and a link-budget perspective, especially in challenging scenarios with sparse infrastructure.

Multi-Connectivity (MC) has emerged as a key mechanism to enhance robustness in such environments by allowing User Equipment (UE) to exploit multiple independent access paths simultaneously. The most widely adopted MC strategy is Full Duplication (FD), also known as selection diversity, in which each packet is transmitted redundantly over multiple interfaces \cite{segura2024empirical, elias2023multi, mishra2025enhanced, mishra2022performance}. This approach is effective in suppressing extreme latency outliers and packet loss, and it is explicitly supported in 3GPP specifications for Ultra-Reliable Low-Latency Communications (URLLC) \cite{3GPP_TS_37_340_R17}. However, unconditional duplication incurs substantial overhead, doubling resource usage regardless of channel conditions, and may therefore be inefficient or impractical in spectrum- and cost-constrained services.

Alternative MC strategies aim to mitigate this overhead. Link aggregation distributes traffic across multiple interfaces without duplication, potentially improving spectral efficiency by lowering the data rate per link and relaxing SINR requirements \cite{kouchaki2020experiment}. Switching-based diversity selects a single active interface at a time and dynamically migrates traffic to an alternative link when degradation is detected \cite{ramirez2025terrestrial, zhang2025empirical}. While theoretically attractive, the effectiveness of both aggregation and switching critically depends on the degree of spatial and temporal independence between links and the temporal granularity and reliability of the switching decision metric. Reference Signal Received Power (RSRP), Signal-to-Interference-plus-Noise Ratio (SINR) or Round Trip Time (RTT) have been typically used as Key Performance Indicator (KPI) to capture the quality of the channel \cite{ramirez2025terrestrial, zhang2025empirical}. Consequently, although aggregation and switching provide resource-efficient alternatives to full duplication, their practical performance in real-world 5G UL conditions depends strongly on link asymmetry, correlation between access paths, reaction delay, and the predictive capability of the selected KPIs in power-limited regimes.

Despite the central role of MC in 5G standardization and its relevance for URLLC, empirical comparisons of duplication, aggregation, and adaptive failover strategies in operational commercial networks remain scarce. Many prior studies rely on simulations \cite{segura2022dynamic, mishra2021multi, ghasemi2025multi} or controlled laboratory environments \cite{segura2024empirical}. There remains a need for systematic field measurements that jointly analyze radio-layer behavior, UL Tx Pwr control, and per-packet E2E latency under realistic mobility and traffic conditions in 5G.

This paper addresses these gaps through extensive real-world measurement campaigns over two commercial 5G Non-Standalone (NSA) networks operated by different Mobile Network Operators (MNOs). Measurements are conducted across urban, suburban, and rural scenarios, with per-packet latency, packet loss, radio KPIs, and UL Tx Pwr control behavior jointly analyzed. The results from the first experiment reveal that rural deployments exhibit fundamentally different behavior, characterized by power-limited UL transmission, severe latency outliers, and limited benefits from aggregation or simple switching strategies. These observations motivate a second experiment focused exclusively on the rural scenario, where the impact of UL data rate reduction, Uplink Transmission Power (UL Tx Pwr) and RSRP are examined in detail.

Building on these insights, this work proposes a Primary-Anchored Adaptive Failover (PAAF) framework that selectively exploits MC through switching or Partial Duplication (PD). Unlike static approaches like FD, PAAF enables the UE to activate redundancy dynamically based on radio and latency indicators, explicitly accounting for service cost and resource utilization. This design aims to approximate the reliability of FD while substantially reducing duplication overhead and the associated operational cost. The proposed mechanisms are evaluated using the collected measurement traces, enabling a direct comparison between FD, aggregation, switching, and adaptive PD under identical real-world conditions.

The main contributions of this paper are as follows:

\begin{itemize}

\item A large-scale empirical characterization of latency, packet loss, and UL Tx Pwr-limited behavior across urban, suburban, and rural 5G deployments under controlled and matched measurement conditions.

\item A control-aware evaluation of multi-connectivity strategies, explicitly modeling KPI observation delay and reaction quantization effects for switching and PD.

\item A probabilistic interpretation of duplication-based reliability under partially correlated rural fading, explaining the structural tail-latency suppression observed in the measurements.

\item A detailed experimental investigation of the relationship between target UL data rate, transmit power, and latency in rural environments, highlighting the limitations of link aggregation under power-limited operation.

\item The design and validation of the PAAF framework, demonstrating that radio- and latency-triggered partial duplication can achieve near-full-duplication reliability with substantially reduced overhead in the evaluated rural scenario.

\end{itemize}

The remainder of this paper is organized as follows. Section~\ref{Sec:Related_Works} reviews the most relevant studies on MC, duplication, switching and link aggregation, highlighting the predominant focus on duplication-based strategies and the limited availability of empirical analyses of bandwidth-splitting schemes in operational 5G networks. Section~\ref{Sec:System_Design} describes the overall system design and methodology, including the functional architecture, operational modes, measurement setup, and performance indicators. Section~\ref{Sec:Experiments} details the measurement campaign conducted across urban, suburban, and rural scenarios, outlining the experimental design, controlled variables, and evaluation methodology. Section~\ref{sec:paaf_algo} introduces the PAAF framework and the decision logic underlying switching and PD -mechanisms, including the discrete-time control-loop model and KPI-triggered activation rules. Section~\ref{Sec:results} presents and analyzes the experimental results, with particular emphasis on uplink latency behavior, packet loss, power-limited operation, and the reliability–overhead trade-off of the evaluated MC strategies. Section~\ref{Sec:Discussion} discusses the main findings, practical implications, and limitations of the study. Finally, Section~\ref{Sec:Conclusion_Future_Work} summarizes the conclusions and outlines future research directions, including adaptive MC mechanisms that dynamically combine PD and load balancing.

\section{Related Works} \label{Sec:Related_Works}

MC has been progressively standardized as a key architectural enhancement to improve reliability, throughput, and mobility robustness in cellular systems. Initially introduced in LTE under the concept of Dual Connectivity (DC), MC enables a UE to maintain simultaneous connections with multiple base stations, typically referred to as Master and Secondary nodes. This architecture was further generalized in 5G New Radio (NR), where MC supports LTE–NR \cite{yilmaz2019overview} and NR–NR \cite{mishra2021multi} combinations and allows for flexible bearer configurations, including split bearers and packet duplication at the Packet Data Convergence Protocol (PDCP) layer \cite{3GPP_TS_37_340_R17}. From a functional perspective, MC provides path diversity by exploiting independent radio links that may operate on different frequency bands, carriers, or even radio access technologies. Different mechanisms for MC are intended to mitigate block errors and transient outages without relying exclusively on retransmission procedures \cite{segura2022dynamic}. Beyond reliability, MC also aims to enhance throughput through parallel resource utilization and to improve mobility performance by reducing handover interruption times \cite{ghasemi2025multi}. Nevertheless, while the standardization framework defines the operational principles and signaling procedures for MC, it does not prescribe optimal strategies for redundancy activation, traffic splitting, or adaptive link selection under realistic radio conditions.

Among the different MC configurations, PD has emerged as the most extensively investigated mechanism for enhancing reliability in 5G systems. At the PDCP layer, duplication allows identical data packets to be transmitted simultaneously over multiple independent radio links, thereby increasing the probability that at least one copy is successfully delivered within the required latency bound \cite{3GPP_TS_37_340_R17}. This approach has been widely advocated for critical services, where stringent packet error rate targets and tight latency constraints demand mitigation of short-lived impairments without relying solely on Hybrid Automatic Repeat reQuest (HARQ) retransmissions. Analytical studies and experimental evaluations have demonstrated that duplication effectively suppresses extreme latency outliers and reduces packet loss, particularly under intermittent fading, mobility-induced degradation, or bursty interference conditions \cite{segura2024empirical, ding2021understanding, majamaa2023enhancing, kaneko2023investigating}. By exploiting spatial, frequency, or operator diversity, duplication transforms independent link failures into a diversity gain that significantly tightens the upper tail of the latency distribution. However, these benefits come at the cost of doubled spectral usage and increased power consumption, thereby raising concerns about efficiency and associated operational cost, as its static application may not represent the most resource- or cost-efficient strategy for heterogeneous scenarios.

In contrast to PD, link aggregation seeks to improve performance by distributing traffic across multiple interfaces without replicating packets \cite{paramita2024flow}. Aggregation can be realized through split bearers at the PDCP layer, where data flows are partitioned between Master and Secondary nodes, enabling concurrent utilization of independent radio resources \cite{elias2023multi}. From an information-theoretic perspective, dividing the offered load across parallel links reduces the required data rate per interface, potentially relaxing the associated SINR constraints and improving robustness under degraded radio conditions. Several analytical and simulation-based studies have explored optimal traffic splitting and resource allocation policies in multi-connectivity environments, showing that aggregation can increase throughput efficiency while maintaining acceptable latency under idealized assumptions of independent fading and flexible scheduler coordination \cite{kuaban2021performance, pupiales2022fast}. Nevertheless, the majority of existing investigations rely on synthetic traffic models, controlled channel conditions, or idealized independence between links. 

As another alternative within the MC design space, switching diversity, also referred to as link selection, seeks to exploit path diversity without incurring the overhead associated with FD. In this strategy, the UE transmits over a single active interface at any given time, while periodically monitoring the quality of multiple candidate links and dynamically switching to the most favorable one when performance degradation is detected. From a theoretical standpoint, switching diversity can achieve diversity gains when link impairments are sufficiently uncorrelated, particularly in heterogeneous deployments involving different operators, frequency bands, or radio access technologies. Key performance indicators (KPIs) such as RSRP, SINR, or RTT are commonly used to trigger switching decisions \cite{ramirez2025terrestrial, zhang2025empirical}, yet the interaction between metric sampling rate, decision latency, link asymmetry, and correlated fading remains insufficiently characterized in real 5G uplink deployments.

Measurement-based studies of MC in rural and heterogeneous environments have recently gained attention, including terrestrial–satellite integration scenarios \cite{lopez2022empirical, lopez2023connecting, ramirez2025multi, zhang2025empirical}. These works provide valuable empirical characterization of throughput, latency, and reliability improvements under duplication-based schemes. In particular, \cite{zhang2025empirical} presents one of the most comprehensive experimental evaluations of 5G MC in remote-control scenarios using terrestrial (5G NSA) and satellite links, demonstrating significant suppression of latency spikes under full duplication across rural and urban deployments. 

However, empirical comparisons between duplication, aggregation, and adaptive PD strategies under controlled and matched dual-operator rural conditions remain scarce. Moreover, the interaction between uplink power-limited operation, infrastructure co-location, link asymmetry, and switching decision-loop dynamics has received limited systematic investigation.

The present work builds directly on the experimental framework of~\cite{zhang2025empirical}, employing synchronized timestamping, dual-operator 5G links, and repeated trajectories to ensure data consistency. This study addresses this gap through a controlled, repeatable field campaign over matched rural trajectories using dual-operator 5G links. Rather than focusing solely on throughput or latency statistics, the analysis emphasizes the structural mechanisms governing diversity effectiveness. Specifically, it examines (i) the impact of infrastructure co-location and correlated fading on achievable diversity gain, (ii) the discrete-time control-loop behavior of switching decisions, (iii) the suppression of latency tail events under PD, and (iv) the interaction between rate-dependent SINR requirements and UL Tx Pwr-control dynamics. By integrating empirical measurements with mechanism-oriented analysis, the present work extends prior campaign-based evaluations toward a system-level understanding of reliability–efficiency trade-offs in MC.

\section{System Design and Methodology} \label{Sec:System_Design}

This section describes the overall system architecture, traffic generation, and the performance indicators used to evaluate and compare the behavior of the different connectivity methods and methodologies adopted in this study.

\subsection{System Architecture}

\begin{figure}[!h]
    \centering
    \includegraphics[width=\columnwidth]{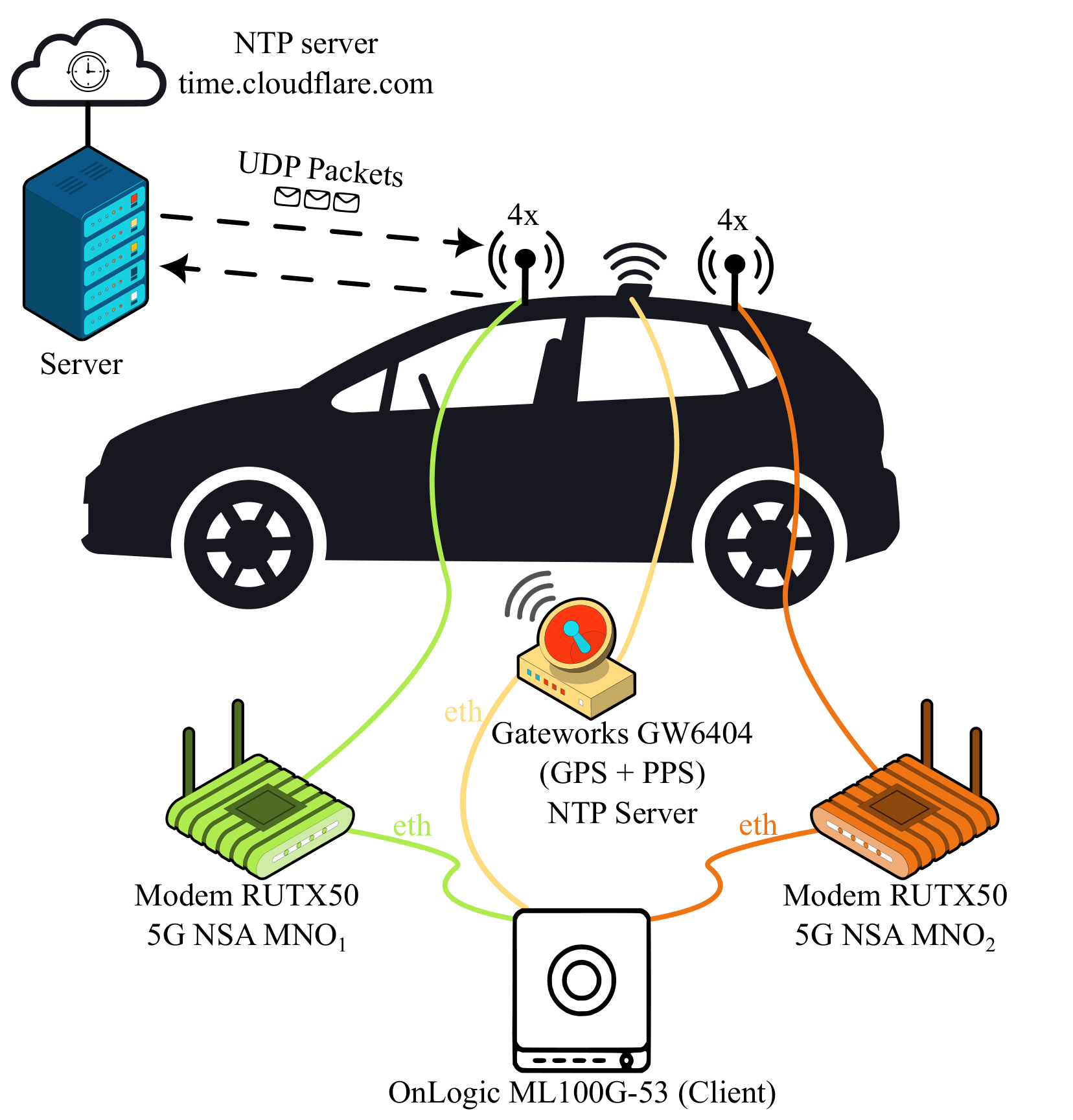}
    \caption{Measurement-system architecture used for synchronized dual-operator 5G NSA data collection}
    \label{fig:setup}
\end{figure}

The architecture of the experimental platform is illustrated in Fig.~\ref{fig:setup}. The core measurement system was installed inside a vehicle and included an OnLogic ML100G-53 industrial computer with an 11th Gen Intel\textsuperscript{\textregistered} Core\textsuperscript{\texttrademark} i3-1115G4 CPU 
and 8~GB of RAM.
Two Teltonika RUTX50 routers were mounted inside the vehicle and connected to the OnLogic computer via Ethernet, each configured with a different SIM card corresponding to one of the MNOs that act as independent communication interfaces, allowing simultaneous data transmission through two distinct 5G networks. To improve signal reception during mobile measurements, eight outdoor 5G antennas are mounted on the roof of the vehicle, as shown in Fig. \ref{fig:setup}. These antennas enable reception across sub-6 GHz bands used by MNO1 and MNO2 for 5G connectivity \cite{zhang2025empirical}. 

In order to achieve accurate timing synchronization on the moving vehicle, a Gateworks GW6404 single-board computer, featuring 
an u-blox ZOE-M8 GPS module, is employed to obtain GPS time combined with a Pulse Per Second (PPS) signal. The Gateworks device acts as a local Network Time Protocol (NTP) server, providing sub-microsecond accuracy. The OnLogic computer synchronizes its system clock via NTP over Ethernet from the Gateworks board, ensuring precise packet timestamping during transmission. Meanwhile, the remote server 
maintains its clock synchronization using the public NTP service \texttt{time.cloudflare.com} providing sub-millisecond accuracy. Therefore, the effective end-to-end (E2E) synchronization accuracy is bounded by the least accurate component. 
This setup ensures consistent and highly accurate time alignment between the OnLogic computer and the remote server, which enables precise calculation of one-way and round-trip latency. The resulting clock offset uncertainty is therefore negligible compared to the latency metrics observed in measurements (in terms of milliseconds), ensuring reliable one-way latency estimation. All network flows, including data streams, NTP synchronization, and control traffic, operate simultaneously during the experiments, ensuring synchronized operation of both measurement interfaces under realistic network conditions.

To power all devices during the measurement campaigns, an external battery is used (48~V, 41.6~Ah, 1996.8~Wh) from WS Technicals, together with a 48 V to 220-240~V, 50~Hz DC-to-AC power inverter from RS PRO.

\subsection{UDP Telemetry and Traffic Generation} \label{subsec:traffic}

To characterize network and radio conditions in parallel with latency and packet loss measurements, the mobile node transmits two types of UDP traffic: (i) a constant-bit-rate measurement stream, configurable in the range of 0.25–4 Mbps, used to compute E2E Key Performance Indicators (KPIs) such as UL and DL latency and packet loss, and (ii) a low-bandwidth telemetry stream that encapsulates radio and positioning metadata for each 5G MNO. Once the radio and positioning data are captured, they are embedded within the payload of the high-rate stream.

The device generates constant-bit-rate UDP traffic with payloads of 1436 Bytes to prevent IP fragmentation. The inter-packet interval in the UL varies depending on the experiment. In the DL, traffic is not independently generated; instead, the server echoes each successfully received UL packet, preserving the payload and updating the transmit timestamp. This mechanism enables per-packet DL one-way latency estimation under symmetric traffic conditions. Each packet contains a flag identifying the transmitting MNO, an 8-byte sequence number and an 8-byte transmit timestamp. The remaining bytes of the payload contain either telemetry information or constant-pattern filler ensuring uniform packet structure. The receiver logs both the arrival timestamp and the sequence number, enabling per-packet computation of one-way latency and packet loss.

Once per second (1~Hz), the device inserts a JSON-encoded telemetry snapshot into the payload a JSON-encoded telemetry snapshot for each MNO. The snapshot aggregates (a) GPS coordinates, (b) modem operational status reflecting connectivity and transmission conditions, (c) serving-cell metrics for LTE (anchor) and NR NSA, and (d) neighboring-cell metrics for LTE. These metadata provide full visibility into the radio environment and support correlation of link-layer conditions with network performance. Combining the per-packet measurement logs with the 1~Hz telemetry enables correlation between bursts of latency and variations in radio metrics. 
In addition, the system incorporates a reverse-path mechanism that routes packets back to the transmitter after reception, enabling consistent E2E evaluation and direct comparison between UL and DL behavior.


\subsection{Key Performance Indicators (KPIs)}

In order to characterize the suitability of commercial 5G NSA networks for remote video monitoring and teleoperation, a set of KPIs that capture both radio-link conditions and E2E service performance are monitored. The selected KPIs are: RSRP, UL Tx Pwr, number of handovers, data rate, end-to-end latency, and application-level packet loss ratio. In the following, a definition of each metric is given to motivate its relevance for the considered experiments.

\textbf{Reference Signal Received Power (RSRP):} is obtained from the cellular modem via AT commands at a sampling period of 1 s. It quantifies the average power of the LTE / 5G NSA reference signals and is a standard indicator of the large-scale radio-channel quality. For teleoperation, RSRP is a key explanatory variable because deep fades and coverage holes lead to reduced link margin and more aggressive link adaptation. 

\textbf{Uplink Transmission Power (UL Tx Pwr):} reports the maximum UE transmit power observed over the previous 1~s interval (1~Hz) \cite{quectel2020rg50xq}. The power reported by the modem reflects how aggressively the UE needs to transmit in order to satisfy the gNB’s Power-Control (PC) target on the Physical UL Shared Channel (PUSCH). In 4G/5G systems, UL PC combines an open-loop component with an optional closed-loop correction \cite{3gpp_ts_38_213}. In simplified form, the UE transmit power per PUSCH transmission can be expressed as

\begin{equation}
P_{\text{Tx,UL}} =
\min\left\{
  \begin{array}{l}
  P_{\max,\text{UE}},\\[2pt]
  P_0 + \alpha PL + \Delta_{\text{UE}} + \mathrm{TPC} + 10\log(2^\mu M)
  \end{array}
\right\},
\label{eq:tx_power}
\end{equation}

\noindent where $P_{\max,UE}$ [dBm] is the maximum transmit power given by the UE (typically 23 dBm), $P_0$ [dBm] is the target power spectral density configured by the network, $\alpha$ is the path-loss compensation factor, $PL$ [dB] is the estimated path loss, $\Delta_{UE}$ [dB] accounts for Modulation and Coding Scheme (MCS) and throughput-dependent offsets, Transmit Power Control (TPC) [dB] is a closed-loop power-control correction term, and $\mu$ is the SCS configuration and $M$ is the bandwidth of the PUSCH expressed in number of physical resource blocks (PRBs).

In the open-loop part of PC, the UE estimates the DL path loss from the RSRP and determines the required $P_{Tx,UL}$ using the configured ($P_0, \alpha$) parameters \cite{3gpp_ts_38_213}. 
In the closed-loop part, the gNB periodically measures the received SINR and issues TPC commands to fine-tune the UE power around the target. 
This mechanism aims to keep the UL SINR within an operating region that balances reliability and inter-cell interference. Nevertheless, when the required power exceeds $P_{\max,UE}$, the UE becomes power-limited and no further compensation is possible for increasing path loss. This situation typically occurs at large distances from the gNB, under strong shadowing, or when higher-order modulation and coding schemes are used, and it is more frequent in rural scenarios where cells are large and gNBs are sparsely deployed. 

\textbf{Number of handovers (HO):} is obtained by counting cell-ID changes reported by the modem. This metric is normalized by the duration (minutes) and distance (kms) of the route. Frequent handovers are known to introduce short service interruptions and to interact negatively with buffer-bloat and scheduling in overloaded cells. Handover bursts along the trajectory can therefore manifest as repeated control-loop glitches.

\textbf{Data rate:} defines the operating point of the UL telemetry stream. Traffic is generated as defined in Section III-\ref{subsec:traffic}. The selected target data rates (4, 2, 1, 0.5, and 0.25 Mbps) were chosen to reflect realistic UL operating points for remote video monitoring and teleoperation applications. A bitrate of 4 Mbps corresponds to a typical low-latency 1080p video stream encoded with H.264/H.265 under moderate compression \cite{black2024evaluation}. Lower rates (2 and 1 Mbps) emulate adaptive bitrate degradation to reduced visual quality or increased compression under constrained radio conditions, while 0.5 and 0.25 Mbps represent highly degraded fallback modes suitable for minimal situational awareness or telemetry-dominant operation. These operating points therefore allow controlled evaluation of the trade-off between video quality, uplink power-limited behavior, and latency reliability in coverage-constrained rural deployments. For each target data rate, the packet inter-sending time is adjusted accordingly. 

Although this is an input parameter rather than a performance outcome, it is treated as a KPI because it directly affects latency and the probability of congesting the link under adverse radio conditions. 

\textbf{E2E latency:} is defined as the one-way propagation time between the transmission of a packet and its reception. Each packet carries a transmit timestamp and a unique sequence number, allowing precise computation of latency as the difference between reception and transmission times after clock synchronization. Both UL and DL latencies are measured independently. The analysis considers the full latency distribution, including the 90th, 95th, and 99th percentiles of the one-way latency. RTT is primarily used to sanity-check clock synchronization.

\textbf{Packet loss ratio:} is defined as an application-level reliability KPI that accounts for both missing packets and packets arriving after the application deadline. The metric is calculated separately for UL and DL directions by comparing transmitted and received sequence numbers. This KPI directly reflects the reliability of the communication link and the effectiveness of redundancy mechanisms in maintaining stable connectivity. A packet is considered useful if it arrives within 800~ms of its transmission time; packets arriving later are marked as “late loss”, as such delays are unacceptable for teleoperation even if the packet is eventually delivered. Packets that do not arrive within a 10~s timeout window are treated as true loss, and for statistical purposes they are assigned a synthetic latency value of 10~s.

\section{Scenarios and Experiments} \label{Sec:Experiments}

This section describes the measurement scenarios and the two experiments conducted to investigate how commercial 5G NSA networks support remote video monitoring and teleoperation. All measurements were performed with two modems mounted in the same vehicle, each connected to a different MNO (MNO1 and MNO2), so that both networks were probed simultaneously under identical mobility and traffic conditions.

\subsection{Scenarios}

\begin{figure*}[!t]
    \centering
    \subfloat[\textbf{Urban scenario (Aalborg city center)}]{%
        \includegraphics[height=0.28\textwidth]{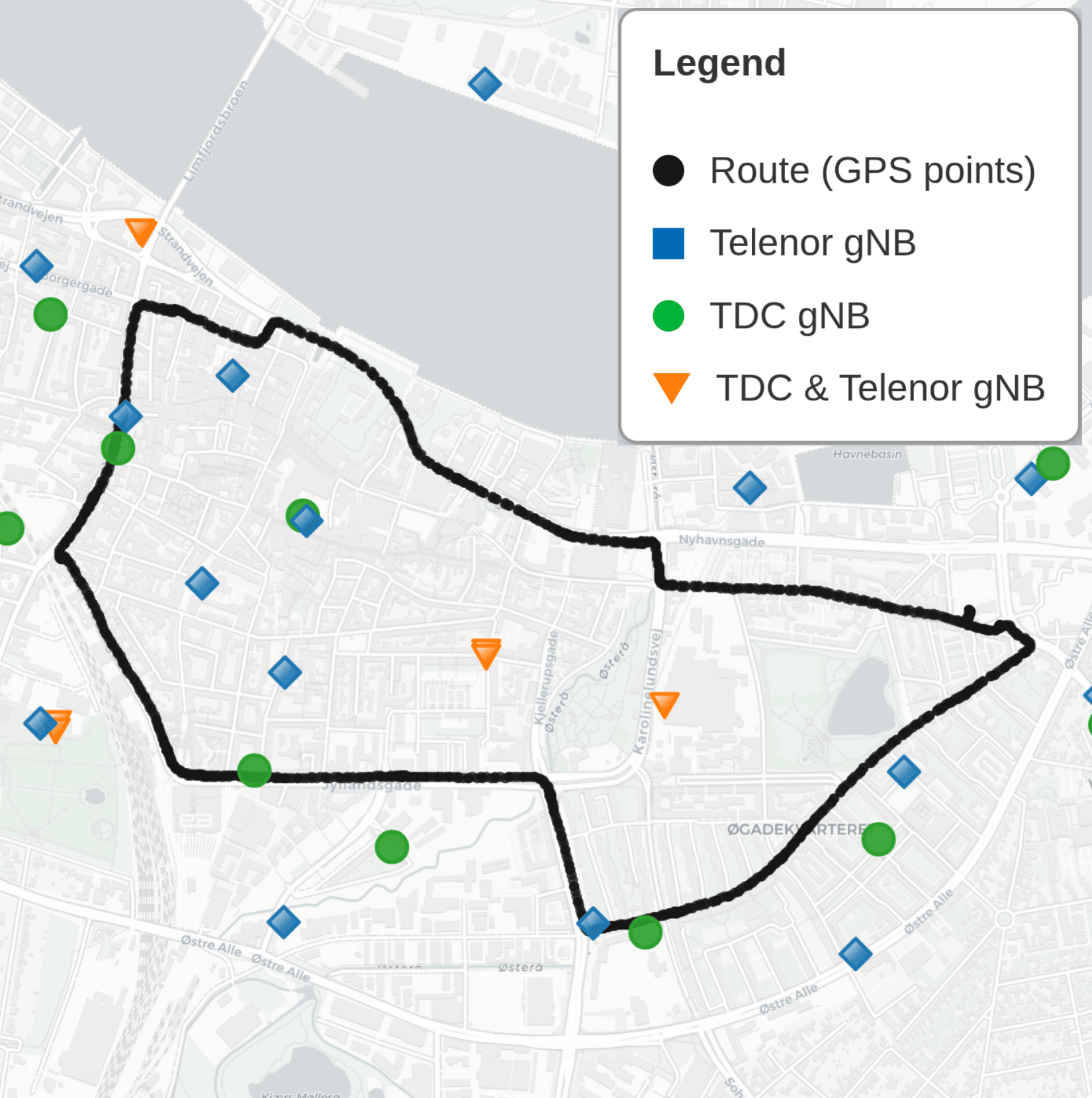}%
        \label{fig:urban_map}}
    \hfil
    \subfloat[\textbf{Suburban scenario}]{%
        \includegraphics[height=0.28\textwidth]{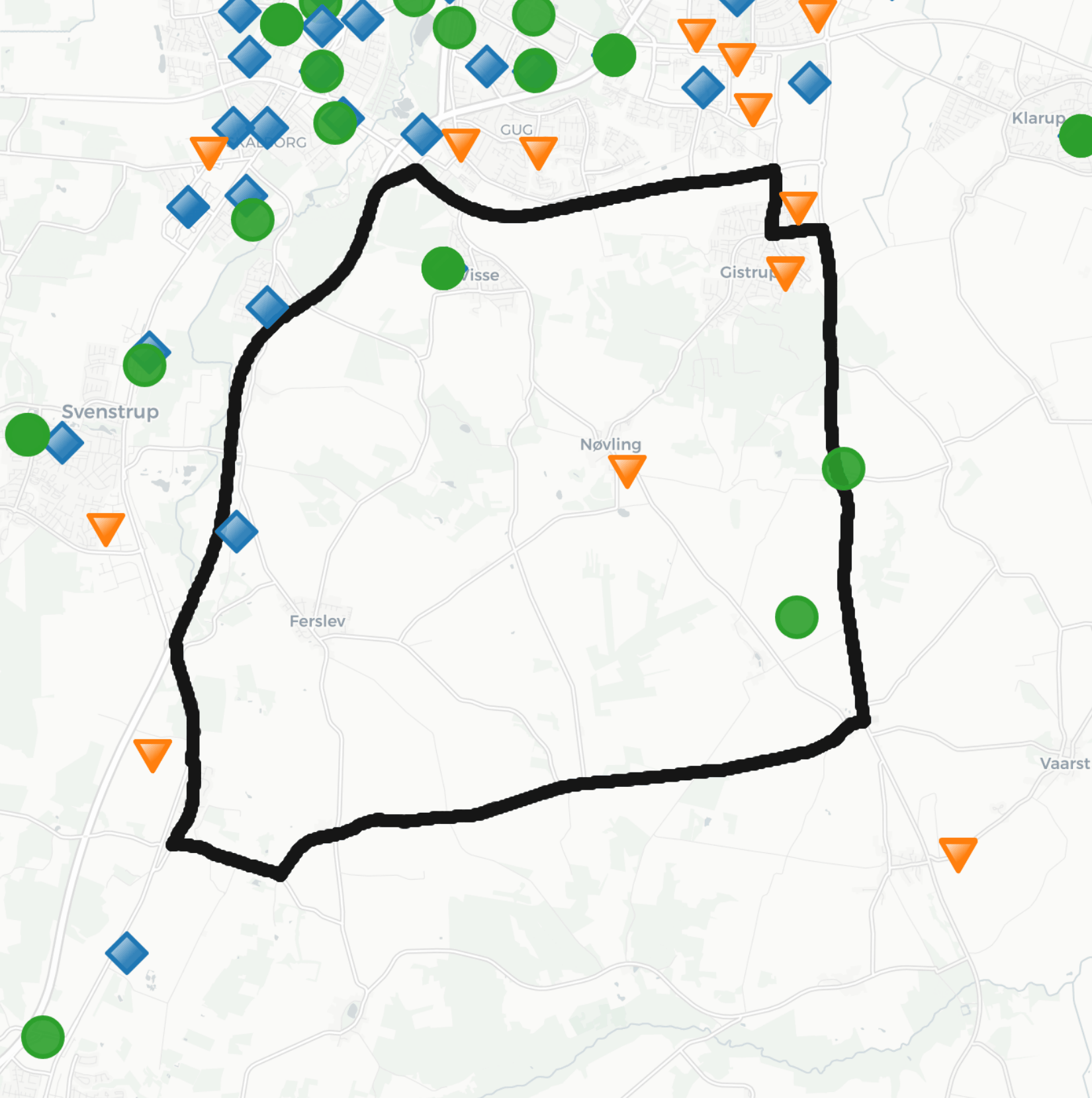}%
        \label{fig:suburban_map}}
    \hfil
    \subfloat[\textbf{Rural scenario}]{%
        \includegraphics[height=0.28\textwidth]{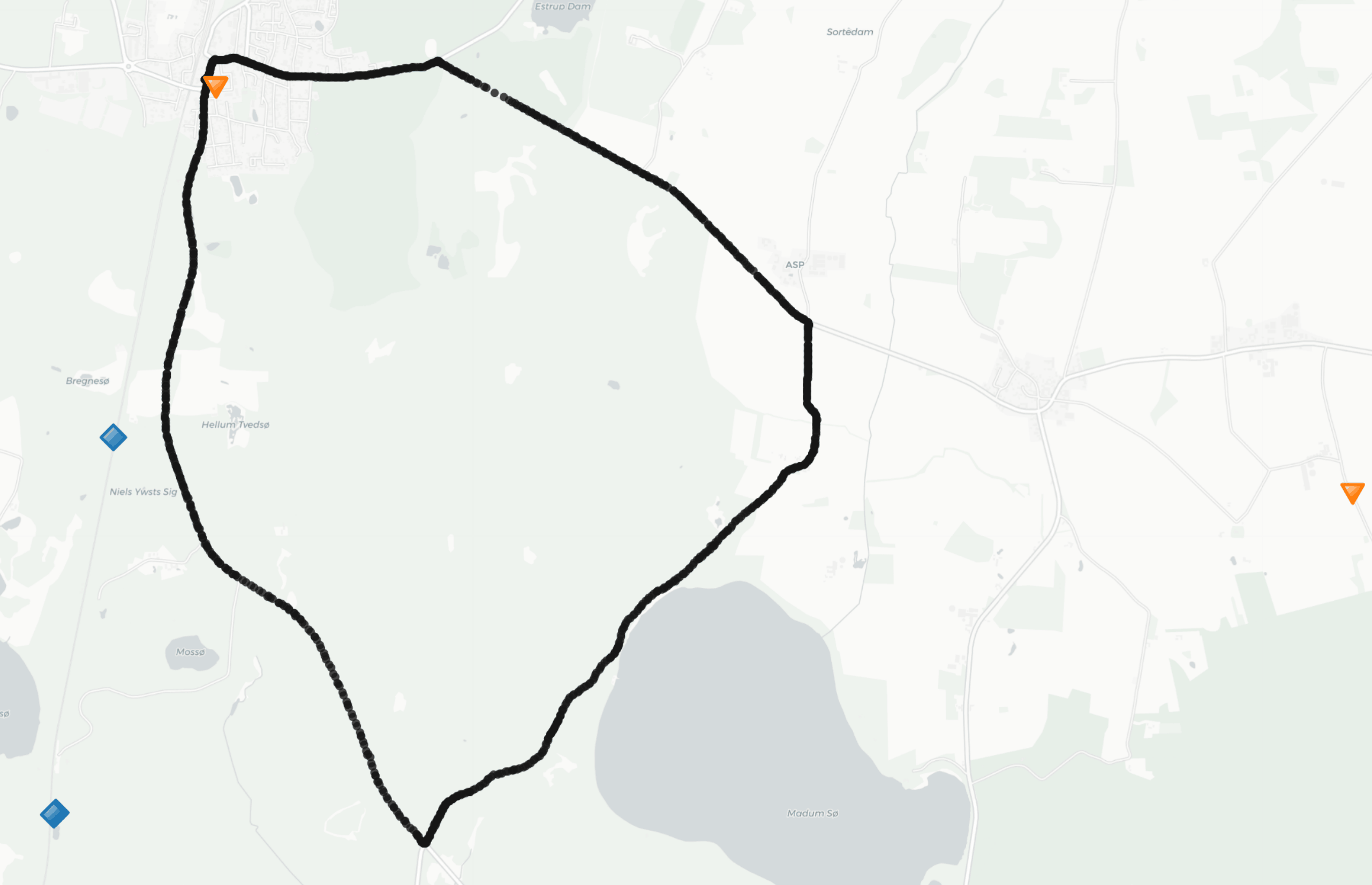}%
        \label{fig:rural_map}}
    \caption{Geographical overview of the measurement routes across the three evaluated scenarios.}
    \label{fig:scenarios}
\end{figure*}

The measurement campaign covers three typical deployment environments: urban, suburban, and rural. Two experiments were conducted, and each route was driven multiple times according to each data-rate. Cruise control was used whenever feasible to maintain an approximately constant vehicle speed and reduce variability caused by driver behavior. These scenarios were chosen to capture a wide range of propagation characteristics and infrastructure layouts typical of commercial 5G deployments in Denmark.

\textbf{Urban scenario:}
The urban environment exhibits high cell density and strong, overlapping coverage from both operators. The route follows a 5~km loop (lasting approximately 18 minutes) through dense residential and commercial areas within the city of Aalborg, providing stable connectivity and lower average latency. The urban route traverses dense city streets with multi-storey buildings, frequent intersections, and heterogeneous traffic conditions. Base stations are deployed with relatively short inter-site distances, and Line-of-Sight (LoS) to at least one site is often available, although intermittent non-LoS conditions occur due to building blockage and street canyons. Buildings and other structures introduce moderate multipath propagation, yet the frequent handovers and overlapping cells may ensure sustained throughput and consistent radio conditions. In such an environment, handover frequency is typically high, but the average RSRP and link margin are expected to be favorable. The urban measurements thus provide a baseline where latency and packet loss are mainly influenced by mobility-induced handovers rather than by coverage limitations.

\textbf{Suburban scenario:} 
The suburban route follows arterial roads and residential areas at the outskirts of the city, with a mixture of open spaces, low-rise buildings, and sporadic industrial zones. It combines sections of both strong and weak coverage, incorporating transitions between suburban and semi-rural areas. The route follows a 32~km loop (lasting approximately 32 minutes). Inter-site distances are larger than in the urban case and building density is lower, leading to smoother large-scale fading but more pronounced transitions between well-covered and weak-coverage areas. 

\textbf{Rural scenario:}  
The rural environment is characterized by sparse base station density, large inter-site distances, and partially obstructed LoS conditions due to vegetation and terrain elevation. In such areas, gNBs from different operators are often co-located on the same towers, reducing spatial diversity. This scenario represents the most challenging case for multi-connectivity, as signal degradation or fading tends to affect both links simultaneously. The rural route follows a 13~km loop (lasting approximately 16 minutes of continuous packet transmission) consisting of open farmland and small residential clusters.

Table \ref{tab:scenarios} summarizes the duration and the distance of each scenario.

\begin{table}[h!]
\centering
\begin{tabular}{lcc}
\hline \hline
\textbf{Scenario} & \textbf{Duration [minutes]} & \textbf{Distance [km]} \\ \hline \hline
Urban    & 18 & 5.2  \\ \hline
Suburban & 32 & 32.1 \\ \hline
Rural    & 16 & 12.4 \\ \hline
\end{tabular}
\caption{Approximate mean duration and travelled distance for each measurement scenario}
\label{tab:scenarios}
\end{table}

Fig.~\ref{fig:scenarios} provides a geographical overview of the three measurement routes. Each map depicts the trajectory of the measurement vehicle (highlighted in black) under distinct coverage conditions: (a) the urban scenario, representing the Aalborg city center with dense cell deployment and strong 5G signal; (b) the suburban scenario, which traverses a suburban transition area combining regions of both high and low coverage; and (c) the rural scenario, where sparse base-station deployment and terrain variations create challenging radio conditions. In the Figs \ref{fig:urban_map} -- \ref{fig:rural_map}, the gNBs of MNO1 (blue squares), MNO2 (green circles) or both (orange triangles)  are also geolocated and placed near the evaluated routes.

\subsection{Experiment 1: Multi-Scenario Evaluation}
\label{subsec:experiment1}

The first experiment aims at characterizing the joint behavior of radio KPIs and E2E performance across all three scenarios under representative teleoperation traffic loads. For each operator, the constant-bit-rate UDP telemetry configuration defined in Section III-C is applied, employing UL target rates of 4, 2, and 1 Mbps.

For each scenario and data rate combination, the vehicle completes the full route while the dual-operator setup records all the KPIs described in previous section. The three rates are tested in immediate succession along the same route, so that the large-scale radio conditions are as similar as possible across configurations. This allows isolating the impact of the offered load on latency and reliability from the impact of spatial variations in coverage.

Experiment 1 has two objectives. First, it provides a comparative baseline of urban, suburban, and rural performance under identical traffic loads for both operators. Second, it examines whether latency spikes and packet losses are associated with RSRP and UL Tx Pwr variations.

\subsection{Experiment 2: Rural Scenario with Extended Data-Rate Sweep}
\label{subsec:experiment2}

As detailed in Section VI\ref{subsec:results_expA}, the rural deployment consistently exhibits higher 95th percentile UL latency (Tables \ref{tab:expA_TELENOR} and \ref{tab:expA_TDC}), increased packet loss probability (Fig. \ref{fig:rsrp_vs_ul_tx_pwr}), and pronounced power-limited behavior (Fig. \ref{fig:tx_pwr_scenarios}), confirming it as the most demanding scenario for teleoperation, with frequent coverage holes and episodes where the UE operates close to its maximum transmit power. To better understand the role of the offered load under such adverse conditions, Experiment~2 focuses exclusively on the rural route and extends the set of evaluated target data rates.

A second measurement campaign was conducted using the 4, 2, and 1~Mbps configurations, together with two additional lower data rates: 0.5 and 0.25~Mbps. For each of the five data rates, the same telemetry pattern and packet size were maintained over the measurement campaign, with the inter-sending time adjusted accordingly. 

This experimental design enables a quasi-controlled comparison: for a given geographical segment of the rural route, it can be observed how the one-way latency distribution and packet loss evolve as the data rate is varied, while the underlying radio conditions remain broadly comparable. 

The richer dataset of Experiment~2 is used for evaluating the proposed PAAF algorithm, which exploits radio and latency information to improve resilience under such challenging conditions.

\subsection{Experiment 2Bis: Theoretical Link-Budget Scaling with Target Uplink Data Rate}

The relationship between the target UL data rate and the robustness of the communication link can be further interpreted through classical information-theoretic arguments. According to Shannon’s capacity theorem \cite{luntovskyy20235g}, the maximum capacity $C$ over a band-limited channel with bandwidth 
$BW$ and signal-to-noise ratio SINR is given by:

\begin{equation}
    C \textrm{[bps]} = BW\textrm{[Hz]} \log_2(1+\textrm{SINR}).
    \label{eq:shanon1}
\end{equation}

\noindent Therefore, the data rate $R < C$ must hold, which leads to:

\begin{equation}
    R \textrm{[bps]} \leq BW \textrm{[Hz]} \log_2(1+\textrm{SINR}).
    \label{eq:shanon2}
\end{equation}

\noindent  Then:
\begin{equation}
\Delta \textrm{SINR [dB]} \geq 
10  \log_{10}\!\left( \frac{ 2^{\frac{R_1}{BW}} -1 }{2^{\frac{R_2}{BW}} -1} \right).
\label{eq:shanon3}
\end{equation}

It is of particular interest to assess whether reducing the UL data rate within a constant $BW$ leads to improved latency performance. While channel quality (e.g., SNR) is independent of the offered traffic load, lowering the required throughput relaxes the spectral efficiency constraints and may enable the scheduler to select more robust MCS configurations. In power-limited rural deployments, this can reduce retransmissions and scheduling delays, potentially improving E2E latency. This would suggest that the UE can partially influence the experienced performance through traffic adaptation, rather than relying solely on operator-side configurations.

\section{Primary-Anchored Adaptive Failover (PAAF) Algorithm}
\label{sec:paaf_algo}

Teleoperation, autonomous navigation, and real-time uplink-dominant telemetry applications require sustained low-latency and high-reliability performance, typically with one-way latency targets below 150 ms and sub-percent packet loss \cite{black2024evaluation, ITU_G114_2003}. In such settings, multi-connectivity can mitigate performance outages, yet FD may incur substantial overhead that is directly translated into a bigger service cost, motivating lightweight controllers that exploit operator diversity without replicating all traffic. 

This work evaluates two PAAF variants designed to support the considered latency and reliability requirements. The first dynamically selects the most suitable interface based on real-time performance indicators, whereas the second activates packet duplication when the primary link operates under degraded conditions. In the first algorithm, PAAF Switching, at any given time, exactly one operator/interface carries the traffic to reduce the overhead, whereas both candidate links are continuously monitored via parallel measurements and telemetry. The second algorithm, PAAF PD, duplicates packets over a secondary link when the primary link operates in a coverage-limited regime, characterized by low RSRP and UL Tx Pwr approaching the UE maximum. This mechanism preserves E2E performance with the tradeoff of a moderate increase in overhead, which is directly proportional to the transmitted throughput and the associated operational cost.

For PAAF Switching, the controller is policy-anchored to a preferred operator $p$ (e.g., contractually or cost-driven) and performs temporary failover to a secondary operator $q$ only when the preferred link violates a performance threshold and the alternative is comparatively better.

The dual-operator setup used in the measurement campaigns naturally enables multi-connectivity, as two independent 5G NSA links are available in parallel. This setting allows us to design and systematically compare a switching policy, which we evaluate offline using the collected traces. The proposed algorithm is implemented at the transport layer and is application-agnostic, yet it exploits both radio-level and network-level KPIs, namely RSRP, UL Tx Pwr, and one-way UL latency. 
Accordingly, the controller combines: (i) an open-loop component, which detects performance degradation from radio conditions isolated or combined, and (ii) a closed-loop component, which reacts to one-way latency measurements observed at the server also isolated or combined with radio KPIs. The dataset of Experiment~2 is the one used to evaluate the effectiveness of this approach under challenging coverage radio conditions.

\subsection{Inputs and Control-Loop Timing Model and Reaction Delay} \label{subsec:time_modelling}

The PAAF controller runs at the UE and has simultaneous visibility of the two 5G links.
\begin{itemize}
    \item \textbf{Radio KPIs (open-loop predictors):} RSRP and UL Tx Pwr sampled at 1~Hz per interface.
    \item \textbf{Latency (closed-loop indicator):} per-packet one-way latency. For a real application with switching algorithm, the UE would send only one packet through the secondary interface to monitor the state of the secondary link.
\end{itemize}

Let $r_i(t)$ denote RSRP (dBm), $u_i(t)$ UL Tx Pwr (dBm), and $\ell_i(t)$ one-way UL latency (ms) for operator $i\in\{p,q\}$.

\subsubsection{Switching Control Timing}

Switching decisions are evaluated periodically at discrete instants
\begin{equation}
    t_k = k T_s, \quad T_s = 100~\text{ms}.    
\end{equation}

At each decision instant $t_k$, the controller evaluates the degradation condition of the primary link and compares it with the secondary interface. If the configured threshold conditions are satisfied at $t_k$, the switching action is triggered immediately at that instant.

If degradation occurs at time $t \in [t_k, t_{k+1})$, the controller can only react at the next evaluation instant. Therefore, the switching reaction time ($T_{\text{react}}^{(\text{SW})}$) is bounded by
\begin{equation}
    T_{\text{react}}^{(\text{SW})} \le T_{\text{obs}} + T_s,
\end{equation}

\noindent where $T_{\text{obs}}$ denotes the observation interval associated with the triggering KPI, which may correspond either to the radio-layer window $T_{obs}^{(R)}$ or to the latency evaluation window $T_{obs}^{(L)}$, depending on the selected control metric.

\subsubsection{Partial Duplication Timing}

In contrast to switching, PD is event-driven with respect to degradation detection on the primary link. Whenever the degradation threshold is exceeded and detected, duplication mode is activated without waiting for a switching evaluation cycle.

Thus, PD does not incur additional quantization delay due to periodic interface selection; its reaction time is dominated by KPI observation latency.

\subsubsection{KPI Observation Delay}

Radio KPIs (RSRP and UL Tx Pwr) are sampled periodically by the modem. While these measurements avoid control-plane feedback delay, they are still constrained by their sampling granularity of 1~Hz. The observation delay for radio-based triggering can therefore be approximated as $T_{obs}^{(R)} \approx 1s$. When latency is used as a KPI, degradation must first be observed through E2E RTT feedback. Let $\textrm{RTT}_p(t)$ and $\textrm{RTT}_q(t)$ denote the RTTs of the primary and secondary interfaces. The observation delay for latency-based triggering can be approximated as:

\begin{equation}
    T_{obs}^{(L)} \approx \min\{\textrm{RTT}_p(t), \textrm{RTT}_q(t)\}.
\end{equation}

\noindent Consequently, latency-based mechanisms inherently experience larger reaction delays in rural scenarios where RTT values are elevated. 

The controller distinguishes between radio-based and latency-based triggering mechanisms, which exhibit fundamentally different reaction characteristics. Due to the 1~Hz reporting granularity of radio KPIs, radio-triggered activation primarily mitigates sustained degradation episodes rather than very short sub-second fading spikes. However, in coverage-constrained or power-limited operation, many latency violations are associated with prolonged retransmission bursts and coverage transitions, for which this granularity remains sufficient.

\subsection{Metric-specific switching policies}
PAAF Switching can be instantiated using a single metric. In all cases, the decision rule is asymmetric: it prioritizes staying on $p$ unless $p$ is degraded, and it returns to $p$ as soon as $p$ recovers or becomes no worse than $q$. Degradation detection is based on threshold criteria applied to radio and performance metrics. The selected thresholds are grounded in practical radio and service considerations. An RSRP level of $\theta_R =$ -100~dBm typically corresponds to coverage-limited operation in macro-cell deployments \cite{3gpp_38133}, where link margin becomes reduced and performance degradation is likely. The uplink transmit power threshold $\theta_U =$ 21~dBm is set close to the nominal maximum UE transmit power (23 dB in NR FR1), thereby identifying near-saturation, power-limited regimes \cite{3gpp_ts_38_213}. Finally, the latency threshold $\theta_L =$ 150~ms reflects the upper bound commonly tolerated in interactive teleoperation systems previously defined in this Section. These thresholds define the operating points beyond which the primary link is considered degraded. Different switching policies are described as follows:

\textbf{RSRP-based failover: }
if the UE is attached to $p$ and $r_p(t)<\theta_R$, the controller checks whether $q$ offers higher RSRP.


\textbf{UL Tx Pwr-based failover: }
high UL Tx Pwr is a direct symptom of a stressed link budget and power-limited operation and $u_p(t)>\theta_U$. 

\textbf{Latency-based failover: }
directly captures end-to-end service quality and can be updated at sub-second granularity under bounded delay regimes. However, as latency-based detection relies on the reception of packets, excessively large delay excursions inherently increase the observability lag, thereby extending the effective switching reaction time.

To enable joint evaluation of the previously defined failover policies (RSRP+UL Tx Pwr and RSRP+UL Tx Pwr+Latency), normalized ``excess'' $e$ terms are defined as follows:
\begin{equation}
e_R(i,t) = \max\!\bigl(0,\theta_R - r_i(t)\bigr),
\end{equation}

\begin{equation}
e_U(i,t) = \max\!\bigl(0,u_i(t) - \theta_U\bigr),
\end{equation}

\begin{equation}
e_L(i,t) = \max\!\bigl(0,\ell_i(t) - \theta_L\bigr).
\end{equation}

A joint score is then computed as:
\begin{equation}
S_i(t)=
w_R\frac{e_R(i,t)}{\sigma_R}+
w_U\frac{e_U(i,t)}{\sigma_U}+
w_L\frac{e_L(i,t)}{\sigma_L},
\end{equation}
where $\sigma_{R,U,L}$ are scaling constants (e.g., $\sigma_R = 5$ dB, $\sigma_U = 2$ dB and $\sigma_L = 100$~ms) and $w_{R,U,L}$ are weights.

\begin{itemize}
    \item \textbf{RSRP + UL Tx Pwr:} set $w_L=0$.
    \item \textbf{RSRP + UL Tx Pwr + Latency:} the weights were empirically optimized using the rural dataset to balance latency-tail suppression and duplication overhead. The selected values ($w_R=0.8$, $w_U=0.7$, $w_L=1.0$) correspond to a stable operating region where tail-latency reduction is achieved without excessive redundancy activation.
\end{itemize}

In all cases, a primary-preferred selection policy was applied, whereby traffic remains anchored to the primary link unless the evaluated risk metric indicates a strictly better alternative path. The decision rule is defined as follows:

\begin{itemize}
    \item If attached to $p$ and $S_p(t)>0$ and $S_q(t)<S_p(t)$, switch to $q$.
    \item If attached to $q$ and $(S_p(t)=0)$ or $S_p(t)\le S_q(t)$, return to $p$.
\end{itemize}

This policy enforces primary anchoring while allowing failover only when the secondary link exhibits a strictly lower risk indicator. At any given time, only one interface remains active.

The resulting behavior can be interpreted as a two-tier control loop. The open-loop tier (via $r_i(t)$ and $u_i(t)$) steers traffic away from links that are likely to become power-limited or coverage-limited, whereas the closed-loop tier (via $\ell_i(t)$) corrects decisions based on actual E2E performance observed.

\subsection{Metric-specific partial duplication policies}
The PAAF PD decision logic is intentionally lightweight, if the primary link exceeds the thresholds for one or various selected KPIs, it starts duplicating. When the primary link becomes stable again and the conditions are below the threshold or thresholds, then it stops duplicating. For the experiment, we have different combinations such as decisions based only on RSRP, UL Tx Pwr or Latency or combination of different ones in \textbf{AND} or \textbf{OR} combinations that means both KPIs MUST exceed the threshold or, at least, one of them.

Because comparative switching or PD can induce ping-pong under marginal differences, we introduce a minimum dwell time before re-switching or start/stop duplicating: 100ms if includes the latency parameter, 1s if not. This preserves the preferred-operator anchoring while preventing pathological oscillations under near-ties.

\subsection{Offline Evaluation Methodology}
The PAAF algorithm is evaluated offline using the rural traces from Experiment~2. Since both operators were measured simultaneously along identical trajectories and with matched data rates, the per-link KPI time series can be replayed into the controller as if it were operating online. For each decision interval $[t_k, t_{k+1})$, the following procedure is applied:

\begin{itemize}
    \item The per-packet latencies and losses of each interface are used to compute $\ell_i(t)$ and to assess constraint violations;
    
    \item Controller reaction delay is explicitly modeled according to the timing analysis introduced Subsection III--\ref{subsec:time_modelling}. In particular:
    \begin{itemize}
        \item Latency-based decisions incur an observation delay 
        $T_{\text{obs}}^{(L)} \approx \min\{\textrm{RTT}_p(t), \textrm{RTT}_q(t)\}$.
        \item Radio-based decisions incur an observation delay 
        $T_{\text{obs}}^{(R)} \approx 1~\text{s}$.
        \item Switching decisions are further quantized by the heartbeat interval $T_s$, such that 
        $T_{\text{react}}^{(\text{SW})} \le T_{\text{obs}} + T_s$.
        \item PD activation is event-driven and therefore 
        $T_{\text{react}}^{(\text{PD})} \approx T_{\text{obs}}$.
    \end{itemize}
    
    \item The 1~Hz RSRP and UL Tx Pwr samples are used to compute $r_i(t)$ and $u_i(t)$.
    
    \item For switching, packets transmitted within the interval $[t_k, t_{k+1})$ are assigned to the interface selected at $t_k$. The resulting path-level latency and packet loss are reconstructed accordingly.
    
    \item For PD, when duplication is active, the effective latency is computed as $\min\{\ell_p(t), \ell_q(t)\}$, and a packet is considered lost only if both interfaces violate the delivery constraint.
    
    \item The overhead of PD is evaluated at the user-plane level as the fraction of packets transmitted over the secondary interface. Control-plane signaling overhead is neglected, as it is negligible compared to duplicated user-plane traffic.
\end{itemize}

Processing time at the UE is negligible relative to observed UL latency (tens to thousands of milliseconds) and is therefore not modeled explicitly. This procedure yields an emulated E2E performance trace for the PAAF strategies, which can be directly compared against single-operator baselines (MNO1 only, MNO2 only) and against purely latency-based switching without radio awareness. Moreover, FD is used as an upper-bound reference for the proposed strategies. The corresponding results, including the fraction of packets below 150~ms, the proportion of late and true loss, and the impact of the heartbeat interval $T_s$, are presented in Section \ref{Sec:results}.

To make the offline evaluation criteria explicit, Table \ref{tab:final_evaluation_metrics} summarizes the final metrics used to compare the evaluated connectivity strategies. These metrics capture both service performance and resource efficiency. Latency percentiles quantify nominal and tail behavior, while late-loss and true-loss ratios distinguish packets that arrive after the application deadline from packets that are not received within the timeout window. Interface usage and duplication overhead quantify the operational cost of switching and partial duplication.

\begin{table}[!h]
\centering
\renewcommand{\arraystretch}{1.15}
\begin{tabular}{p{0.30\columnwidth} p{0.62\columnwidth}}
\hline \hline
\textbf{Metric} & \textbf{Definition and interpretation} \\
\hline \hline
UL latency percentiles &
90th, 95th, 99th, and 99.9th percentiles of one-way uplink latency, used to quantify nominal and tail-latency behavior. \\ \hline

DL latency percentiles &
90th, 95th, and 99th percentiles of one-way downlink latency, used to verify whether degradation is uplink- or downlink-dominated. \\ \hline

Packets below 150 ms &
Fraction of packets with one-way UL latency $\leq 150$ ms, used to measure compliance with the target latency requirement. \\ \hline

Late loss &
Fraction of packets arriving after 800 ms, corresponding to packets delivered too late for teleoperation. \\ \hline

True loss &
Fraction of packets not received within the 10 s timeout, corresponding to hard delivery failures. \\ \hline

Interface usage &
Fraction of packets transmitted over MNO1 and MNO2, indicating switching behavior or primary/secondary interface usage. \\ \hline

Duplication overhead &
Fraction of packets additionally transmitted over the secondary interface, used to quantify redundancy cost for PD and FD. \\ \hline

Normalized overhead/cost &
Duplication overhead weighted by a secondary-link cost factor, used to compare reliability against relative resource or service cost. \\ \hline
\hline
\end{tabular}
\caption{Final evaluation metrics used for strategy comparison.}
\label{tab:final_evaluation_metrics}
\end{table}

The implementation of all evaluated strategies (including baseline, link aggregation, FD, and PAAF Switching and PD) and the post-processed dataset used for the offline trace evaluation are available in \cite{alvarez_5g_code_2026, alvarez_5g_dataset_2026}, respectively. The repository includes the controller logic, trace replay framework, anonymized datasets, and documentation sufficient to reproduce the reported results.

\section{Results} \label{Sec:results}

This section presents the main findings from both experiments described in the previous Section. Experiment~1, which compares urban, suburban, and rural deployments under three representative data rates (4, 2, and 1~Mbps) is analyzed, and then move to Experiment~2, which refines the analysis for the rural scenario for data rates of 4, 2, 1, 0.5, 0.25~Mbps.

\subsection{Experiment 1: Multi-Scenario Performance}
\label{subsec:results_expA}

Tables \ref{tab:expA_TELENOR} and \ref{tab:expA_TDC} summarize the KPIs over the complete routes for the Experiment 1 and for MNO1 and MNO2, respectively. For each combination of scenario, operator, and data rate, the tables report the 5th, 50th, 95th percentiles of RSRP and UL Tx Pwr, including the 99th percentile for UL Tx Pwr, also the total number of handovers, the 90th/95th/99th percentiles of one-way latency, for both UL and DL, and the fractions of packets classified as late loss ($>800$~ms) and true loss (timeout at 10~s).

\begin{table*}[!h]
\centering
\small
\setlength{\tabcolsep}{6pt}
\renewcommand{\arraystretch}{1.05}
\resizebox{0.9\textwidth}{!}{

\begin{tabular}{llccccccccc}
\hline \hline
Scenario    &        & \multicolumn{3}{c}{URBAN} 
                       & \multicolumn{3}{c}{SUBURBAN} 
                       & \multicolumn{3}{c}{RURAL} \\
\cmidrule(lr){3-5} \cmidrule(lr){6-8} \cmidrule(lr){9-11}

Target Mbps &        & 4 & 2 & 1 & 4 & 2 & 1 & 4 & 2 & 1 \\ 
\hline \hline

\multirow{3}{*}{RSRP [dBm]}
  & \textbf{5\%}  & -59 & -59 & -60 & -72 & -69 & -72 & -64 & -70 & -70 \\
  & \textbf{50\%} & -73 & -72 & -73 & -87 & -87 & -89 & -90 & -92 & -91 \\
  & \textbf{95\%} & -90 & -88 & -89 & -109 & -105 & -108 & -109 & -110 & -109 \\
\hline

\multirow{4}{*}{UL Tx Pwr [dBm]}
  & \textbf{5\%}  & -39.0 & -40.0 & -35.0 & -24.0 & -28.0 & -24.0 & -24.0 & -20.0 & -23.0 \\
  & \textbf{50\%} & -16.0 & -17.0 & -15.0 & -2.0  & -3.0  & -2.0  & 18.0  & 16.0  & 15.0 \\
  & \textbf{95\%} & 6.0   & 3.0   & 3.0   & 22.0  & 21.5  & 21.5  & 22.5  & 22.5  & 22.5 \\
  & \textbf{99\%} & 13.0  & 9.0   & 11.0  & 23.0  & 23.0  & 23.0  & 22.5  & 23.0  & 22.5 \\
\hline

Handovers/(km $\times$ min)
  &      & 0.38 & 0.39 & 0.47 & 0.09 & 0.09 & 0.10 & 0.08 & 0.13 & 0.08 \\
\hline

\multirow{3}{*}{UL Latency [ms]}
  & \textbf{90\%} & 39  & 41  & 41  & 51  & 53  & 35  & 309 & 45  & 37  \\
  & \textbf{95\%} & 214 & 70  & 735 & 169 & 336 & 46  & 1332 & 85  & 44  \\
  & \textbf{99\%} & 275 & 440 & 870 & 594 & 793 & 663 & $10^{4}$ & 410 & 184 \\
\hline

\multirow{3}{*}{DL Latency [ms]}
  & \textbf{90\%} & 22  & 22  & 21  & 26 & 23 & 22 & 29 & 29 & 28 \\
  & \textbf{95\%} & 28  & 25  & 44  & 29 & 27 & 24 & 32 & 33 & 32 \\
  & \textbf{99\%} & 166 & 214 & 259 & 76 & 91 & 45 & 50 & 59 & 51 \\
\hline

\multirow{2}{*}{Packet Loss [\%]}
  & \textbf{ $\geq$ 800 ms}      & 0.1 & $2\cdot10^{-2}$ & 3.9 & 0.8 & 1.0 & 0.9 & 7.5 & 0.1 & 0.2 \\
  & \textbf{==}$\mathbf{10^{4}}$ \textbf{ms}  & 0.1 & $2\cdot10^{-2}$ & $6\cdot10^{-2}$ &
  0.6 & 0.6 & 0.3 &
  1.6 & $2\cdot10^{-2}$ & $2\cdot10^{-2}$ \\
\hline

\end{tabular}
}
\caption{Summary of KPIs for MNO1 at 4/2/1Mbps. }
\label{tab:expA_TELENOR}
\end{table*}

\begin{table*}[!h]
\centering
\small
\setlength{\tabcolsep}{6pt}
\renewcommand{\arraystretch}{1.05}
\resizebox{0.9\textwidth}{!}{

\begin{tabular}{llccccccccc}
\hline \hline
Scenario    &        & \multicolumn{3}{c}{URBAN} 
                       & \multicolumn{3}{c}{SUBURBAN} 
                       & \multicolumn{3}{c}{RURAL} \\
\cmidrule(lr){3-5} \cmidrule(lr){6-8} \cmidrule(lr){9-11}

Target Mbps &        & 4 & 2 & 1 & 4 & 2 & 1 & 4 & 2 & 1 \\ 
\hline \hline

\multirow{3}{*}{RSRP [dBm]}
  & \textbf{5\%}  & -60 & -62 & -54 & -62 & -62 & -61 & -64 & -65 & -64 \\
  & \textbf{50\%} & -72 & -75 & -70 & -78 & -76 & -77 & -107 & -108 & -107 \\
  & \textbf{95\%} & -92 & -93 & -82 & -97 & -96 & -96 & -118 & -119 & -118 \\
\hline

\multirow{4}{*}{UL Tx Pwr [dBm]}
  & \textbf{5\%}  & -33.0 & -27.0 & -29.0 & -27.0 & -24.0 & -27.0 & -9.0 & -11.0 & -8.0 \\
  & \textbf{50\%} & -8.0  & -1.0  & -5.0  & 1.0   & 2.0   & -1.0  & 22.5 & 22.5 & 21.5 \\
  & \textbf{95\%} & 14.0  & 19.0  & 10.0  & 21.5  & 21.5  & 21.0  & 23.0 & 23.5 & 23.5 \\
  & \textbf{99\%} & 21.5  & 21.5  & 15.0  & 22.5  & 22.5  & 21.5  & 23.5 & 23.5 & 23.5 \\
\hline

Handovers/(km $\times$ min)
  &      & 0.51 & 0.57 & 0.33 & 0.07 & 0.06 & 0.06 & 0.21 & 0.22 & 0.21 \\
\hline

\multirow{3}{*}{UL Latency [ms]}
  & \textbf{90\%} & 23 & 27 & 26 & 34 & 32 & 32 & $10^{4}$ & $10^{4}$ & 1830 \\
  & \textbf{95\%} & 26 & 37 & 30 & 115 & 82 & 52 & $10^{4}$ & $10^{4}$ & $10^{4}$ \\
  & \textbf{99\%} & 111 & 82 & 40 & 246 & 255 & 405 & $10^{4}$ & $10^{4}$ & $10^{4}$ \\
\hline

\multirow{3}{*}{DL Latency [ms]}
  & \textbf{90\%} & 21 & 21 & 19 & 28  & 24  & 22  & 40  & 35  & 35 \\
  & \textbf{95\%} & 24 & 24 & 21 & 52  & 35  & 30  & 62  & 54  & 54 \\
  & \textbf{99\%} & 60 & 79 & 78 & 187 & 159 & 148 & 133 & 122 & 137 \\
\hline

\multirow{2}{*}{Packet Loss [\%]}
  & \textbf{$\geq$ 800 ms}      & 0.1 & $3\cdot10^{-2}$ & $1\cdot10^{-2}$ & 0.2 & 0.1 & 0.2 & 46.2 & 34.7 & 21.9 \\
  & \textbf{==}$\mathbf{10^{4}}$ \textbf{ms}  & 0.1 & $3\cdot10^{-2}$ & $1\cdot10^{-2}$ & 0.1 & 0.1 & 0.2 & 30.9 & 18.5 & 9.9 \\
\hline
\end{tabular}
}
\caption{Summary of KPIs for MNO2 at 4/2/1Mbps.}
\label{tab:expA_TDC}
\end{table*}

\paragraph*{Urban scenario}

In the urban case, the RSRP distributions of the two operators are similar across all three data rates. Despite these comparable radio conditions, the remaining KPIs exhibit marked operator-specific differences.

First, the UL transmission power reported for MNO2 is consistently higher than for MNO1. For example, the median UL Tx Pwr at 1~Mbps is around $-5$~dBm for MNO2, while it remains below $-15$~dBm for MNO1, and the 95th and 99th percentiles are also shifted upwards for MNO2. Moreover, the normalized number of handovers is different between MNO1 and MNO2, indicating a differently planned cell layout in the urban area as seen on the Fig. \ref{fig:scenarios}.

Second, the latency KPIs show the UL and DL latencies are systematically higher for MNO1 than for MNO2. At 4~Mbps, MNO1 exhibits substantially larger 95th and 99th percentile UL latencies and a noticeable fraction of packets exceeding 800~ms, whereas MNO2 maintains much lower high-percentile values and negligible late loss. The same pattern persists at 2 and 1~Mbps. Thus, in the urban scenario, MNO1 operates with lower transmit power and fewer handovers but experiences higher latency and packet loss than MNO2.

\paragraph*{Suburban scenario}

In the suburban scenario there is a clear asymmetry between operators. Here, RSRP is consistently better for MNO2 than for MNO1: MNO1’s median and 95th percentile RSRP are approximately $10$~dB lower, whereas MNO2 maintains stronger signal levels across all data rates. The UL transmission power percentiles and the normalized handover counts, on the other hand, are of similar magnitude for both operators.

These worse radio conditions for MNO1 translate into higher UL latencies and higher packet loss. For all three data rates, MNO1’s 95th and 99th percentile UL latencies exceed those of MNO2, and the fractions of late and true loss are also larger. In contrast, DL latency behaves differently: MNO2 exhibits higher DL 99th percentiles than MNO1 despite its more favorable RSRP, especially at the higher data rates. Hence, in the suburban case, the operator with better DL radio conditions (MNO2) does not necessarily achieve lower DL latency.

\paragraph*{Rural scenario}

The rural scenario reveals the largest performance differences between the two operators. RSRP is significantly better for MNO1 than for MNO2 across all data rates: MNO2’s median RSRP is approximately 10~dB lower, and its 95th percentile reaches deep into the cell-edge region. This translates directly into UL transmission power: for MNO2, both the median and the upper percentiles of UL Tx Pwr operate very close to $P_{\max,\text{UE}}$, while MNO1 retains some headroom except in the worst segments.

These harsher radio conditions for MNO2 are reflected in the latency and packet-loss metrics. At all three data rates, MNO2 shows severely degraded performance in UL and DL latencies for this scenario, and very high fractions of late and true loss. In contrast, MNO1 still experiences degraded latency and some packet loss in the same scenario, but the high-percentile values remain substantially lower and the fraction of true loss is much smaller. These observations suggest that the MNO1 link retains greater stability along the evaluated rural route compared to MNO2.

\begin{figure}[!h]
    \centering
    \includegraphics[width=\columnwidth]{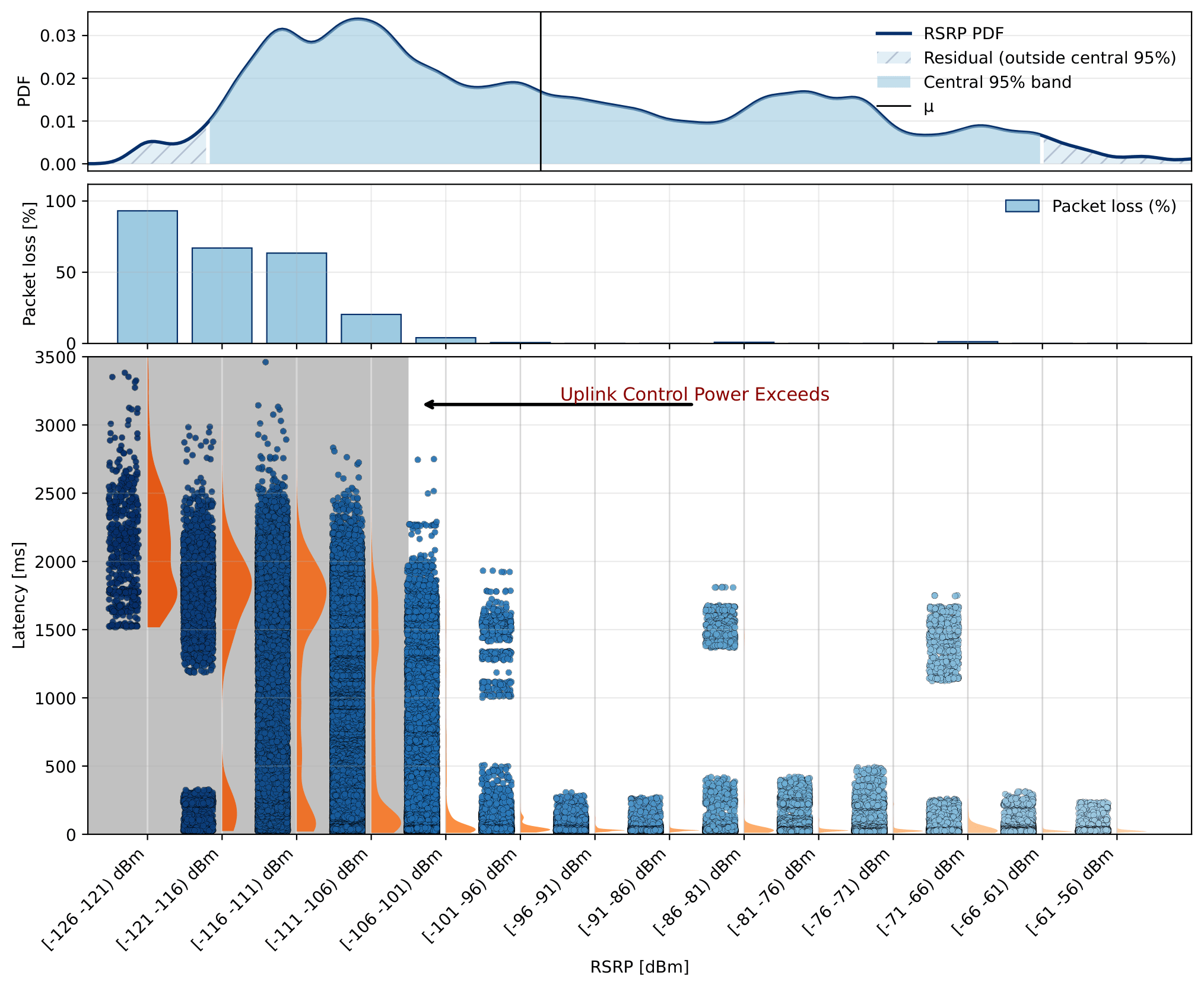}
    \caption{Distribution of UL Latency according to the RSRP level on the UE.}
    \label{fig:rsrp_latency_rural4}
\end{figure}

Fig.~\ref{fig:rsrp_latency_rural4} illustrates the joint distribution of RSRP and UL one-way latency for the rural scenario at 4~Mbps for both operators, with RSRP grouped in 5~dB bins. The bottom panel shows the per-packet latency samples together with their empirical density for each RSRP bin, while the middle panel reports the corresponding packet-loss ratio at that RSRP group level and the top panel the overall RSRP probability density.

A clear step-like behavior is visible around $\mathrm{RSRP}\approx -101$ dBm. For RSRP values above this threshold, the bulk of the latency samples lies well below 500~ms and the distributions are relatively compact, with modest packet-loss ratios. Outliers below this threshold are associated with handovers. Once RSRP falls below $-101$~dBm, both the median and the spread of the latency increase sharply: the distributions become much wider, with a substantial fraction of packets exhibiting delays of several seconds, and the packet-loss ratio rises to very high levels. The dense cloud of points in the shaded low-RSRP region indicates that the UE spends a significant amount of time in conditions that are severely degraded.

This step-like performance transition is consistent with operation in a power-limited regime discussed above: below $-101$~dBm the UL Tx Pwr control frequently saturates at $P_{\max,\text{UE}}$, so additional path loss can no longer be compensated, leading to increased BLER, repeated retransmissions, and consequently to the pronounced latency spikes and packet loss observed in the Fig. \ref{fig:rsrp_latency_rural4}.

\paragraph*{Comparison across scenarios}

Comparing scenarios for each operator confirms the expected pattern: both networks deliver the best performance in the urban route, intermediate performance in the suburban route, and the worst in the rural route due to the distribution of resources on each scenario. However, the magnitude of the degradation differs. For MNO1, the transition from suburban to rural leads to moderate increases in latency and packet loss, whereas for MNO2 the same transition causes orders-of-magnitude increases in the high-percentile latencies and in the late/true loss fractions. The RSRP and UL transmission power statistics clearly show that this is driven by much more severe coverage limitations for MNO2 in the rural environment.

\begin{figure*}[!t]
    \centering
    \subfloat[\textbf{Urban scenario}]{%
        \includegraphics[width=0.33\textwidth]{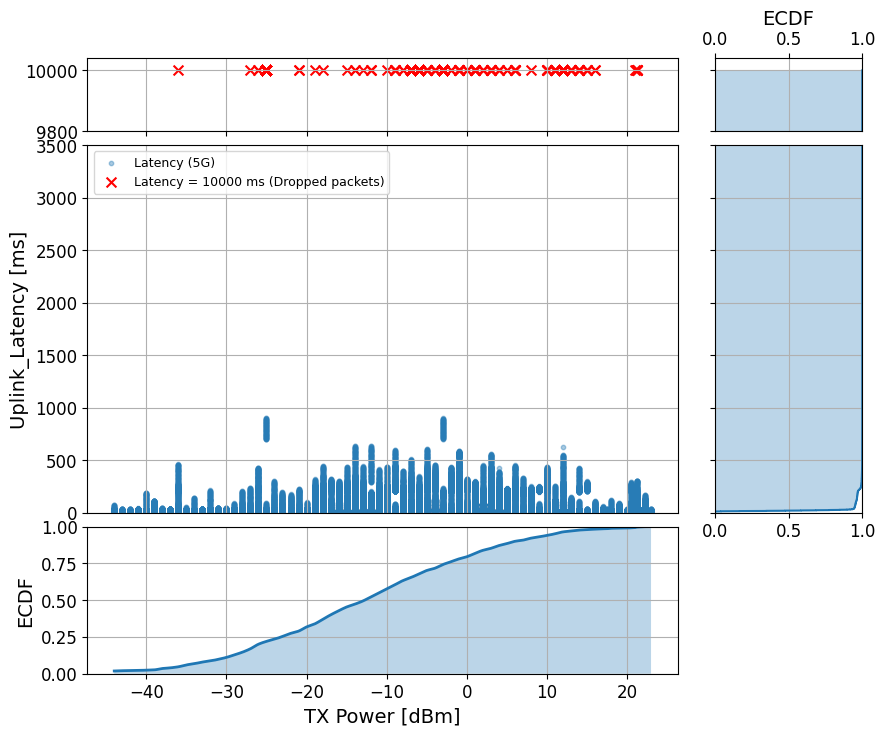}%
        \label{fig:tx_pwr_urban}}
    \hfil
    \subfloat[\textbf{Suburban scenario}]{%
        \includegraphics[width=0.33\textwidth]{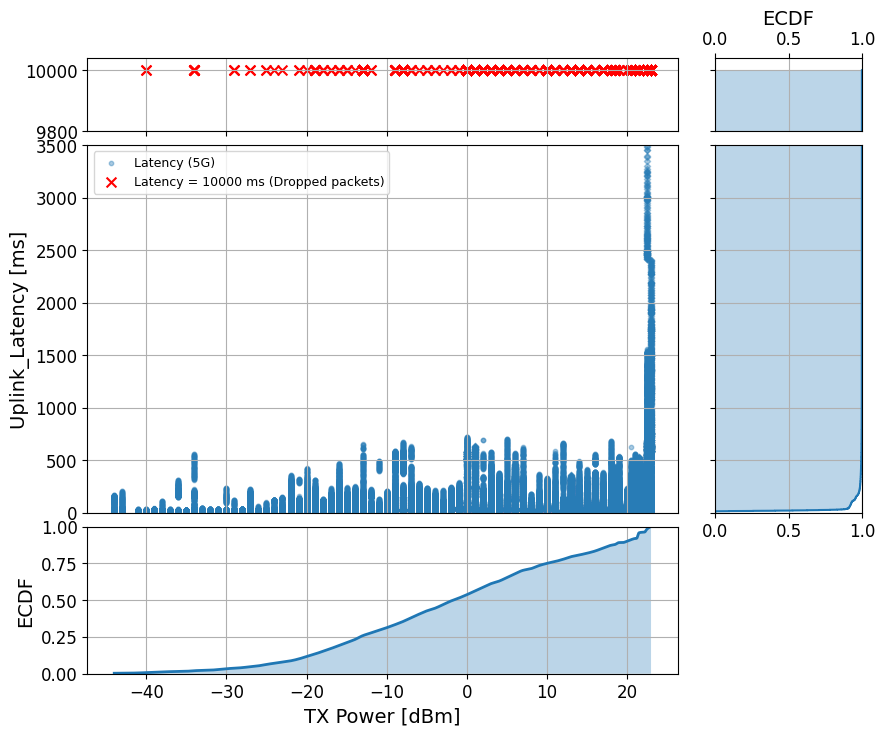}%
        \label{fig:tx_pwr_suburban}}
    \hfil
    \subfloat[\textbf{Rural scenario}]{%
        \includegraphics[width=0.33\textwidth]{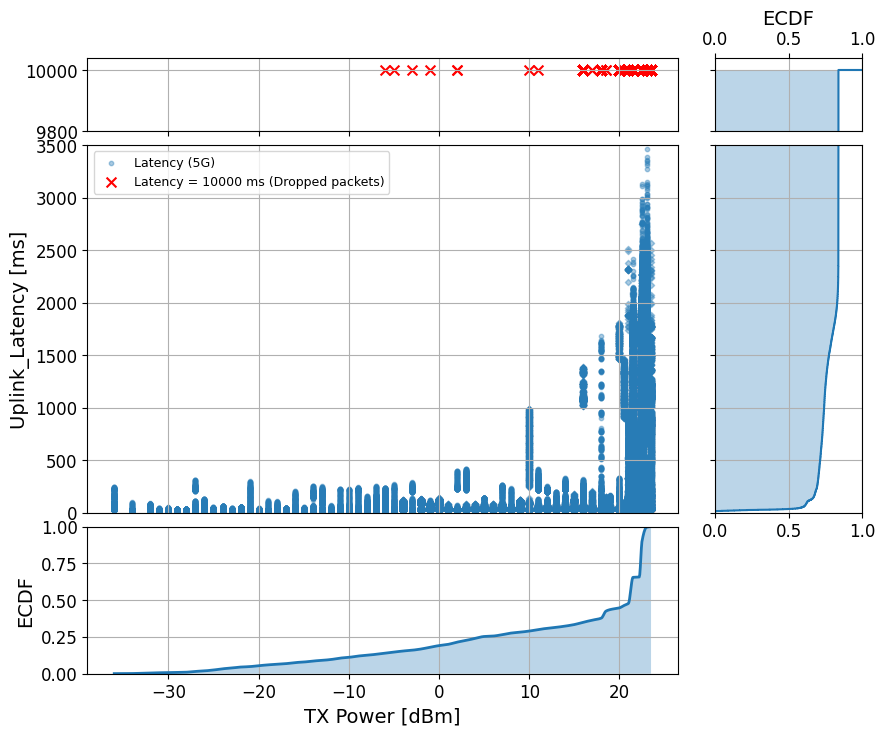}%
        \label{fig:tx_pwr_rural}}
    \caption{Evaluation of the Latency vs the UL Tx Pwr along the scenarios for Experiment 1}
    \label{fig:tx_pwr_scenarios}
\end{figure*}

Fig.~\ref{fig:tx_pwr_scenarios} further explores these differences by relating UL one-way latency to the instantaneous UL Tx Pwr for the three scenarios at 4~Mbps. Each subfigure shows the per-packet latency samples as a function of transmit power, the Empirical Cumulative Density Function (ECDF) of the transmit-power distribution (bottom panel), and the ECDF of the latency distribution (right panel). Packets that did not arrive within the 10~s timeout are represented as red crosses at 10~s, in line with the definition adopted for late and true loss.

In the urban scenario, UL Tx Pwr spans a broad range, but most samples are concentrated between approximately $-20$ and $10$~dBm. Within this range, the majority of latency samples remain below a few hundred milliseconds, and the latency ECDF is steeped, indicating a relatively tight distribution. Dropped packets at 10~s are present but sparse and are not strongly clustered at the highest transmit powers, suggesting that outages are relatively rare and not solely driven by power-limited operation.

In the suburban scenario, a clearer dependence on transmit power appears. For moderate powers, latency remains mostly below 500~ms, but as the UL Tx Pwr approaches the upper end of the range (around 20--23~dBm), the latency cloud becomes noticeably more dispersed and extends towards several seconds. The number of packets mapped to 10~s also increases in this high-power region. The UL transmit-power ECDF shows that only a fraction of samples operate close to $P_{\max,\text{UE}}$, but these samples contribute disproportionately to the latency tail and loss statistics.

Finally, the rural scenario exhibits the strongest coupling between transmit power and latency. A substantial portion of the samples operate at very high transmit powers, with the ECDF indicating that the UE frequently resides close to its maximum power. In this region, the latency distribution becomes extremely wide: many packets experience delays of several seconds, and there is a dense cluster of packets assigned to 10~s, reflecting frequent outages. Compared to the urban and suburban cases, high transmit power in the rural route is thus a much more reliable predictor of severe latency spikes and packet loss, consistent with the presence of extended power-limited segments and coverage holes.

Table~\ref{tab:rsrp_tx_pwr_correlation} reports the correlation between RSRP and the UL Tx Pwr across the target data rates by using Spearman's rank correlation coefficient $\rho$ \cite{spearman1961proof}, which captures monotonic association and is less sensitive to non-Gaussianity, outliers, and saturation effects that are common in field radio measurements. The use of this metrics provides a compact assessment of whether the relationship is predominantly linear or generally monotonic.

\begin{table}[h]
\centering
\begin{tabular}{cc}
\hline \hline

\textbf{Scenario} & \textbf{Spearman} $\rho$ (Mbps(4\textbar2\textbar 1) \\ \hline \hline
URBAN    & -0.61\textbar-0.68\textbar-0.66   \\ \hline

SUBURBAN & -0.72\textbar-0.68\textbar-0.71  \\ \hline

RURAL    & -0.87\textbar-0.88\textbar-0.87  \\ \hline
\end{tabular}
\caption{Correlation of the RSRP and UL Tx Pwr at different scenarios and data rates.}
\label{tab:rsrp_tx_pwr_correlation}
\end{table}

Across all evaluated data rates and scenarios, the correlation coefficients are consistently negative, indicating a stable inverse relationship. Nevertheless, the rural scenario exhibits systematically larger magnitudes, suggesting a stronger coupling under coverage-limited conditions. As the scenario becomes more challenging and the coverage weakens, the RSRP decreases and the UE tends to transmit at higher power, whereas better RSRP is associated with reduced transmit power. This correlation between RSRP and UL Tx Pwr is primarily governed by the UL Tx Pwr-control and link-budget dynamics, rather than by the offered load alone. 

\subsection{Experiment 2: Rural Performance vs. Data Rate}
\label{subsec:results_expB}

\begin{table*}[!t]
\centering
\small
\setlength{\tabcolsep}{6pt}    
\renewcommand{\arraystretch}{1.05}  
\begin{tabular}{llccccc}
\hline \hline
Target Mbps &        & 4 & 2 & 1 & 0.5 & 0.25 \\
\hline \hline

\multirow{4}{*}{Share of samples in UL [\%]}
  & MNO1 $\le$ 150 ms,\ MNO2 $\le$ 150 ms & 78.0 & 88.9 & 93.1 & 94.2 & 97.2 \\
  & MNO1 $>$ 150~ms,\ MNO2 $\le$ 150~ms   &  6.8 &  3.5 &  0.5 &  0.9 &  0.7 \\
  & MNO1 $\le$ 150~ms,\ MNO2 $>$ 150~ms   & 13.7 &  7.7 &  6.4 &  4.9 &  2.1 \\
  & MNO1 $>$ 150~ms,\ MNO2 $>$ 150~ms     &  1.6 &  $4\cdot10^{-3}$ &  0 &  $5\cdot10^{-3}$ &  0 \\
\hline

\multirow{6}{*}{UL Latency [ms]}
  & \textbf{90\%} MNO1 & 85  & 46 & 42 & 43 & 50 \\
  & \textbf{90\%} MNO2     & 457 & 57 & 67  & 48 & 46 \\ \cline{2-7}
  
  & \textbf{95\%} MNO1 & 283  & 74  & 51 & 52 & 56 \\
  & \textbf{95\%} MNO2     & 1941 & 705 & 317  & 142 & 65 \\ \cline{2-7}
  
  & \textbf{99\%} MNO1 & 1952 & 531 & 105 & 123 & 93 \\
  & \textbf{99\%} MNO2 & $10^{4}$ & $10^{4}$ & 1715 & 992 & 459 \\ \cline{2-7}
\hline

\multirow{6}{*}{DL Latency [ms]}
  & \textbf{90\%} MNO1 & 29  & 28 & 24 & 23 & 23 \\
  & \textbf{90\%} MNO2     & 26  & 23 & 22 & 20 & 20 \\ \cline{2-7}
  
  & \textbf{95\%} MNO1 & 35  & 32 & 27 & 26 & 24 \\
  & \textbf{95\%} MNO2     & 37  & 26 & 26 & 22 & 21 \\ \cline{2-7}
  
  & \textbf{99\%} MNO1 & 92 & 63 & 45 & 43 & 41 \\
  & \textbf{99\%} MNO2     & 89 & 48 & 58 & 50 & 37 \\
\hline

\multirow{6}{*}{Packet Loss [\%]}
  & MNO1 $\boldsymbol{\geq}$ $\mathbf{800}$ \textbf{ms} & 2.8  & 0.5 & $5\cdot10^{-2}$ & 0.5 & 0.1 \\
  & MNO2 $\boldsymbol{\geq}$$\mathbf{800}$ \textbf{ms} & 8.8  & 4.9 & 2.7 & 2.5 & 0.7 \\
  & Both IFs $\boldsymbol{\geq}$$\mathbf{800}$ \textbf{ms} & $4\cdot10^{-3}$ & 0 & 0 & 0 & 0 \\ \cline{2-7}
    
  & MNO1 \textbf{==}$\mathbf{10^{4}}$ \textbf{ms} & 0.9 & $8\cdot10^{-2}$ & $4\cdot10^{-2}$ & 0.3 & $2\cdot10^{-2}$ \\
  & MNO2 \textbf{==}$\mathbf{10^{4}}$ \textbf{ms} & 4.7 & 2.1 & 0.9 & 0.2 & 0.3 \\
  & Both IFs \textbf{==}$\mathbf{10^{4}}$ \textbf{ms} & $4\cdot10^{-3}$ & 0 & 0 & 0 & 0 \\ \cline{2-7}
\hline
\hline

\multirow{4}{*}{Full Duplication Latency [ms]}
  & UL Latency [ms] \textbf{95\%} & 53  & 37 & 38 & 38 & 41 \\
  & UL Latency [ms] \textbf{99\%} & 185 & 50 & 79  & 58 & 58 \\ \cline{2-7}

    & Packet Loss [\%] $\boldsymbol{\geq}$ $\mathbf{800}$ \textbf{ms} & $4 \cdot 10^{-3}$ & 0 & 0 & 0 & 0 \\

  & Packet Loss [\%] \textbf{==}$\mathbf{10^{4}}$ \textbf{ms} & $4 \cdot 10^{-3}$ & 0 & 0 & 0 & 0 \\ \cline{2-7}
\hline

\multirow{4}{*}{Link Aggregation [ms]}
  & UL Latency [ms] \textbf{95\%} & 221  & 89 & 63 & 59 & N/A \\
  & UL Latency [ms] \textbf{99\%} & $10^{4}$ & 1511 & 891  & 264 & N/A \\ \cline{2-7}

  & Packet Loss [\%] $\boldsymbol{\geq}$ $\mathbf{800}$ \textbf{ms} & 2.7 & 1.4 & 1.5 & 0.4 & N/A \\

  & Packet Loss [\%] \textbf{==}$\mathbf{10^{4}}$ \textbf{ms} & 1.1 & 0.5 & 0.2  & 0.2 & N/A \\ \cline{2-7}
  
\hline

\end{tabular}
\caption{Summary of Experiment~2 in rural scenario with extended data rates.}
\label{tab:expB_summary}
\end{table*}

Due to the challenging propagation conditions and the correlation of the radio parameters, Experiment~2 focuses exclusively on the rural scenario and extends the set of evaluated UL data rates to 4, 2, 1, 0.5, and 0.25~Mbps. A dedicated measurement campaign was conducted on a different day for this experiment; consequently, the results are not directly comparable in absolute terms to those of Experiment~1, although the same evaluation methodology is applied. The main goal of this experiment is to investigate how progressively lowering the required data rate alters the effective quality of the UL channel, as perceived at the transport and application layers. For each target rate, the vehicle traverses the same rural route with cruise control enabled and under similar weather conditions. Table~\ref{tab:expB_summary} summarizes, for Experiment~2 and each target data rate, the fraction of samples for which the UL latency is $\leq150$~ms on each operator (and jointly), high-percentile one-way UL/DL latency statistics (90/95/99\%), and late/true loss fractions (thresholds at $800$~ms and $10^{4}$~ms), including simultaneous events across both interfaces. In addition, results for FD are included as an upper-bound performance reference. Link Aggregation is also reported, where the target data rate is evenly split across the two interfaces (e.g., 4~Mbps corresponds to 2~Mbps per link), enabling a direct comparison with the single-link baseline and illustrating the performance gains achievable through diversity. Aggregation results at 0.25~Mbps are not available (N/A), as this configuration would require operating each link at 0.125~Mbps, a rate that was not measured during the experimental campaign.

\begin{figure*}[!t]
    \centering
    \subfloat[\textbf{4 Mbps}]{%
        \includegraphics[width=0.33\textwidth]{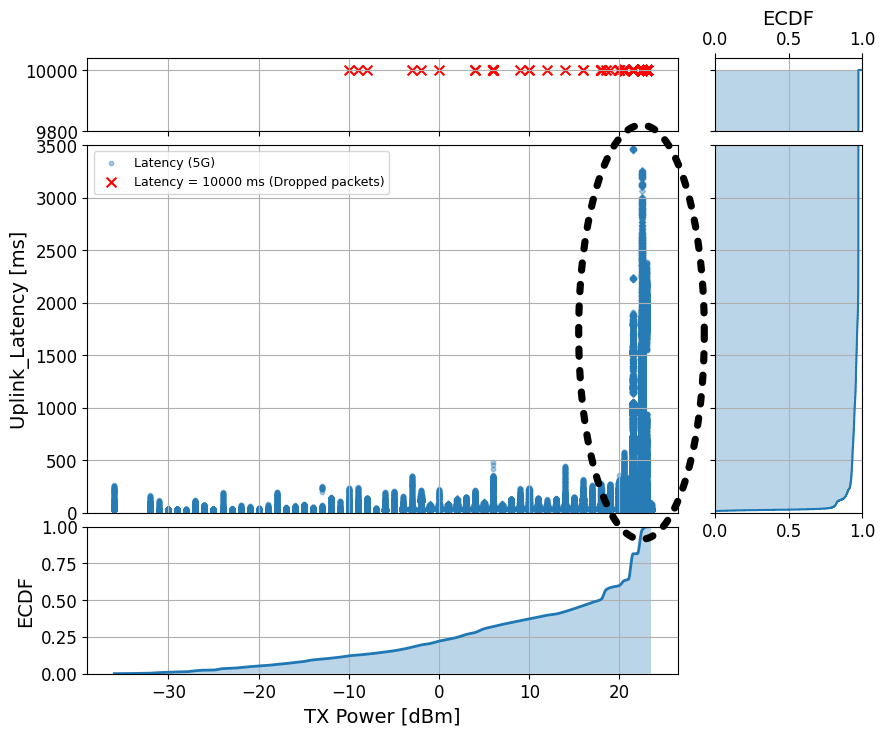}%
        \label{fig:exp2_tx_pwr_rural4}}
    \hfil
    \subfloat[\textbf{1 Mbps}]{%
        \includegraphics[width=0.33\textwidth]{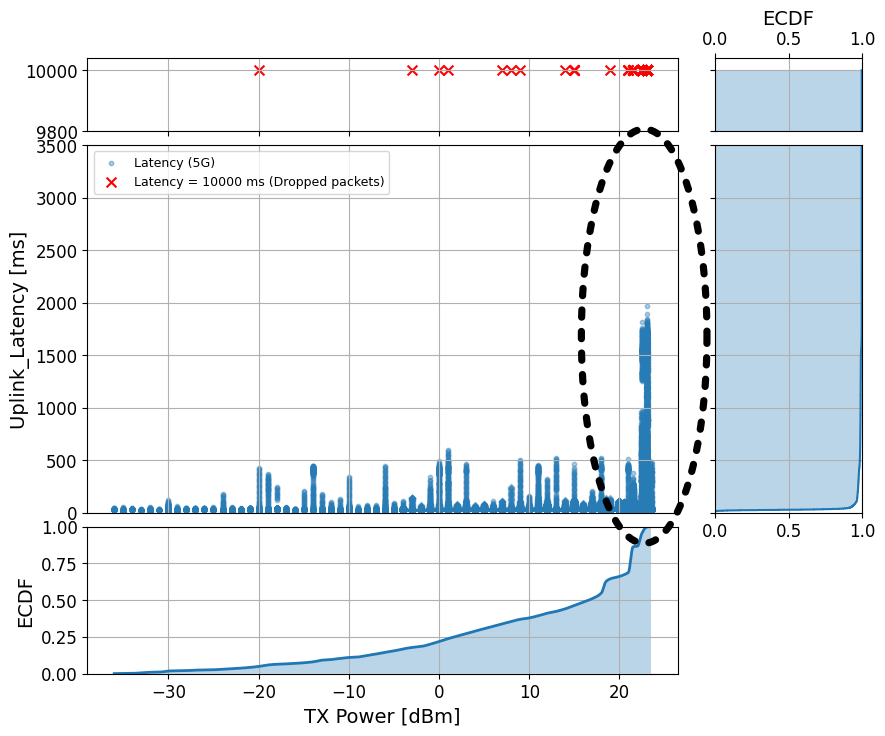}%
        \label{fig:exp2_tx_pwr_rural1}}
    \hfil
    \subfloat[\textbf{0.25 Mbps}]{%
        \includegraphics[width=0.33\textwidth]{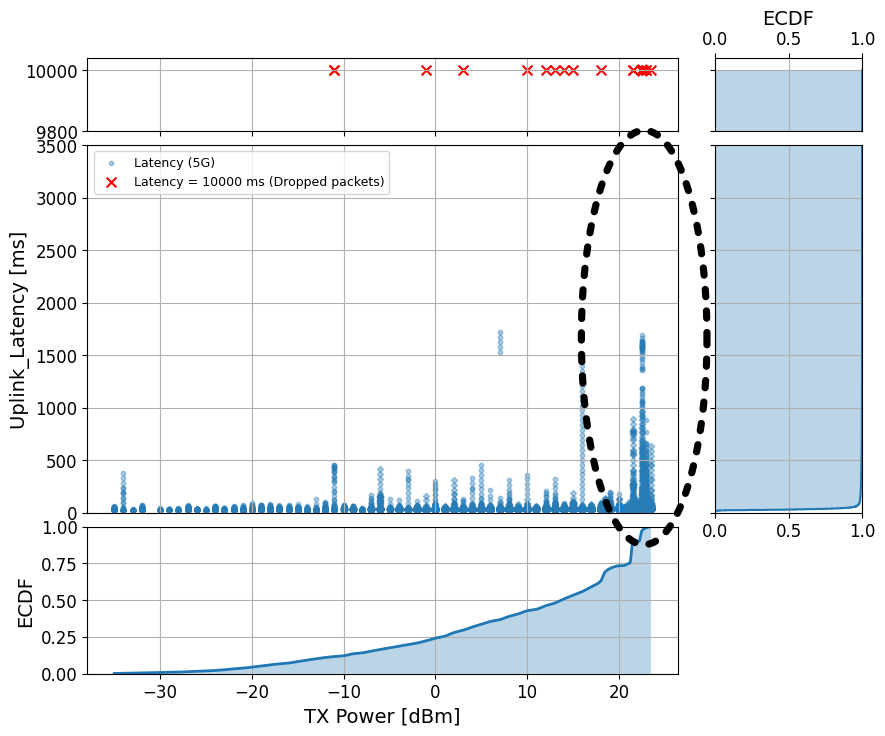}%
        \label{fig:exp2_tx_pwr_rural0_25}}
    \caption{Relationship between UL Tx Pwr and UL latency in the rural scenario for Experiment 2 at target data rates of 4, 1, and 0.25 Mbps. The samples are shown for the evaluated rural trace to illustrate how high UL Tx Pwr values are associated with latency-tail expansion, particularly at higher target data rates.}
    \label{fig:tx_pwr_exp2}
\end{figure*}

Fig.~\ref{fig:tx_pwr_exp2} illustrates the joint distribution of UL Tx Pwr and UL latency in the rural scenario for representative target data rates. he region enclosed by the black dashed circle highlights that, at 4 Mbps, a pronounced latency tail is observed, with a concentration of extreme delays occurring when the UE operates near its maximum transmit power. As the target rate decreases to 1 Mbps and further to 0.25 Mbps, the severity and frequency of these latency outliers progressively diminish. This behavior indicates that reducing the offered load alleviates power-limited operation, thereby mitigating extreme latency excursions while preserving baseline performance. 

The operator-specific statistics in Table \ref{tab:expB_summary} show that this behavior is more severe for MNO2 than for MNO1, as MNO2 exhibits larger high-percentile UL latency and higher late-loss and true-loss ratios at the same target data rate.

Using the UL one-way latency requirement ($\leq150$~ms) as a link-state indicator, the measurements partition into four mutually exclusive regimes: both links compliant, only MNO1 compliant, only MNO2 compliant, and neither link compliant. At higher target data rates, the UL latency distributions exhibit pronounced upper tails, particularly for MNO2, including instances that reach the trace-processing cap used to encode extreme delays/true loss. As the target data rate decreases, the UL latency tails contract for both operators, consistent with fewer excursions into degraded UL states. Correspondingly, reducing the target rate shifts the sample distribution toward the regime in which both links satisfy the UL latency constraint, whereas higher target rates are associated with a greater presence of regimes where at least one link violates the constraint.

In contrast, the DL latency percentiles remain comparatively stable across data rates and between operators, with narrower dispersion and without occurrences of capped extreme-delay events in the DL statistics. This pattern indicates that the dominant variability captured in Table~\ref{tab:expB_summary} is concentrated in the UL path under rural conditions, consistent with UL limitations (e.g., elevated UL Tx Pwr and path loss compensation constraints) rather than DL-side congestion effects. Moreover, the asymmetric UL regimes are not balanced between operators: MNO1 exhibits higher compliance with the UL latency requirement and fewer severe tail events than MNO2, reflecting operator-dependent differences in rural link robustness along the same route.

Finally, the late/true loss indicators (defined via latency thresholds at $>800$~ms and $=10^{4}$~ms, respectively) follow the UL tail behavior and the associated state transitions. Late and true loss events are more prominent at higher target rates and occur more frequently on MNO2 than on MNO1, while simultaneous late/true loss across both interfaces is rarely observed, indicating that extreme impairments are typically not fully synchronized between the two operators along the same trajectory.

\subsection{Data Rate Reduction and Uplink Power-Control Behavior}
\label{subsec:results_channel_expB}

In principle, for a fixed resource allocation (i.e., fixed bandwidth and number of spatial layers), reducing the target UL data rate relaxes the minimum required effective SINR for successful decoding. In practical 5G NR systems, this manifests through link adaptation: for the same allocated resources, a lower transport rate allows the scheduler to select a more robust MCS, which operates at a lower SINR threshold. Consequently, Packet Error Rate (PER) decreases, retransmissions become less frequent, and the link operates in a more stable regime.

Although Shannon’s capacity expression provides useful upper-bound intuition regarding the relationship between rate and required SINR, practical 5G NR systems do not operate in the continuous-capacity regime assumed by (\ref{eq:shanon1})~--~(\ref{eq:shanon3}). Instead, performance is governed by discrete MCS levels, link adaptation, and uplink power-control dynamics. 

To quantify the rate-dependent variation in transmit power, a linear model of the form $\mathrm{UL\ Tx\ Pwr} = a \cdot RSRP + b$ was fitted to the UL Tx Pwr versus RSRP samples for each target rate along the same rural route and under comparable propagation and mobility conditions. To isolate the rate-dependent power adaptation behavior, samples where the UE operated at or near maximum transmit power were excluded. Specifically, observations with $P_{\mathrm{Tx}} \geq 22.5$~dBm were removed to avoid saturation effects, beyond which the UE cannot further increase transmit power to meet the target SINR. During all rural measurements, the MNO2 nominal carrier bandwidth remained constant at 100~MHz, as verified from the logged radio configuration parameters collected during the campaign.

\begin{figure}[!h]
    \centering
    \includegraphics[width=\columnwidth]{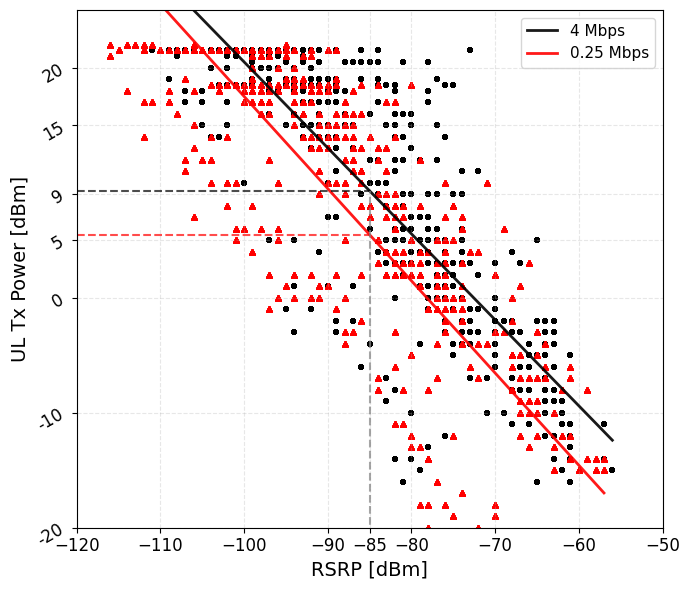}
    \caption{RSRP vs UL Tx Pwr at 4 and 0.25 Mbps}
    \label{fig:rsrp_vs_ul_tx_pwr}
\end{figure}

Fig. \ref{fig:rsrp_vs_ul_tx_pwr} shows the distribution of UL Tx Pwr related to RSRP when reducing from 4~Mbps (black circles) to 0.25~Mbps (red triangles). Evaluating the regression models near the median RSRP value ($-85$~dBm) yields an average transmit-power reduction of approximately 4~dB for a $\times 16$ rate reduction. 

Table~\ref{tab:ul_tx_power_gain} reports the average UL Tx Pwr variation relative to the 4~Mbps baseline for progressively lower target data rates.

\begin{table}[h]
\centering
\begin{tabular}{c c c c}
\hline
\textbf{Target Rate} & \textbf{Rate Reduction} & \textbf{$\Delta$UL Tx Pwr [dBm]} \\
\hline
4 Mbps   & Reference       & N/A \\
2 Mbps   & $\times$2 lower & $1.5$ \\
1 Mbps   & $\times$4 lower & $1.4$ \\
0.5 Mbps & $\times$8 lower  & $2.3$ \\
0.25 Mbps& $\times$16 lower & $3.8$ \\
\hline
\end{tabular}
\caption{Average UL Tx Pwr variation relative to the 4 Mbps baseline for different target data rates.}
\label{tab:ul_tx_power_gain}
\end{table}

From an information-theoretic perspective, under a simplified Shannon-capacity interpretation, reducing the spectral-efficiency requirement can be associated with a lower required SINR. However, the exact reduction is not fixed at 3 dB per halving because it depends on bandwidth, operating point, and practical MCS constraints. However, the empirical results reported in Table~\ref{tab:ul_tx_power_gain} deviate from this theoretical trend. In particular, the observed transmit-power reduction does not scale linearly with the rate reduction, and remains significantly below the expected cumulative gain (e.g., $\approx 12$~dB for a $\times 16$ reduction).

This discrepancy can be attributed to the discrete and system-level nature of practical 5G NR UL operation. Specifically, link adaptation relies on a finite set of MCS, resource allocation is dynamically adjusted by the scheduler, and UL Tx Pwr control follows fractional path-loss compensation rather than ideal SINR targeting. As a result, reductions in the target data rate are often absorbed through changes in allocated bandwidth and coding rate, rather than translating directly into proportional reductions in transmit power. Furthermore, in coverage-limited regimes, the UE frequently operates close to its maximum transmit power, which constrains the achievable power adaptation gain.

The regression parameters for the two extreme operating points are summarized in Table~\ref{tab:regression_params}.


\begin{table}[h]
\centering
\begin{tabular}{c c c}
\hline
\textbf{Target Rate} & \textbf{Slope $a$} & \textbf{Intercept $b$ [dBm]} \\
\hline
4 Mbps   & $-0.75$ & $-54.25$ \\
0.25 Mbps & $-0.80$ & $-62.67$ \\
\hline
\end{tabular}
\caption{Slope intercept gain model for UL Tx Pwr as a function of RSRP}
\label{tab:regression_params}
\end{table}

The slopes of both regressions are similar (approximately $-0.8$), indicating that for each 1~dB degradation in RSRP, the UE increases its transmit power by approximately 0.8~dB. This behavior is consistent with fractional path-loss compensation in uplink power control. As RSRP decreases toward the cell edge, UL Tx Pwr increases proportionally until the maximum transmit-power constraint is reached, beyond which no further power-control compensation is possible and the link operates in a power-limited regime.

The slight difference in slope between the two regression models implies that the transmit-power gap is RSRP-dependent rather than constant. The results indicate that across the measurable RSRP range, the 4~Mbps configuration consistently requires higher transmit power than the 0.25~Mbps configuration under comparable radio conditions. Evaluating the regression models at representative RSRP values illustrates this behavior. At $RSRP = -100$~dBm, the 4~Mbps configuration requires approximately $3$~dB more transmit power than 0.25~Mbps, whereas at $RSRP = -60$~dBm the required difference reduces to approximately $5$~dB. This trend reflects the combined effect of the higher effective SINR requirement associated with the more aggressive MCS and the fractional path-loss compensation applied by UL Tx Pwr control.

\subsection{PAAF results}
\label{subsec:PAAF_algorithm}

This subsection evaluates the performance of the proposed PAAF mechanisms in the rural scenario of Experiment 2 at a target UL data rate of 4~Mbps. The analysis focuses on the ability of PAAF to suppress extreme UL latency and packet loss while minimizing the additional overhead associated with multi-connectivity. Two classes of strategies are considered: PAAF Switching, where traffic is routed over a single interface at any given time, and PAAF PD, where packets are duplicated only when the primary link exhibits signs of degradation. Moreover, a sensitivity analysis was conducted to select the PAAF triggering thresholds..

\subsubsection{Threshold selection via sensitivity analysis}

The rural scenario constitutes the most challenging operating condition, characterized by extended power-limited regimes, correlated fading across operators, and severe UL latency spikes in the single-operator baselines. Consequently, it represents a stringent stress test for adaptive multi-connectivity control.

To instantiate the PAAF decision rules with operationally meaningful parameters, a sensitivity analysis over candidate thresholds for RSRP and UL Tx Pwr was performed. The objective was to identify values that achieve a favorable reliability–efficiency trade-off: (a) suppressing the high-percentile latency tail that dominates service degradations in rural coverage, while (b) keeping the duplication activation (and thus overhead) limited to those segments where the primary link exhibits clear symptoms of link-budget stress.

\begin{figure}[!h]
    \centering
    \includegraphics[width=0.95\columnwidth]{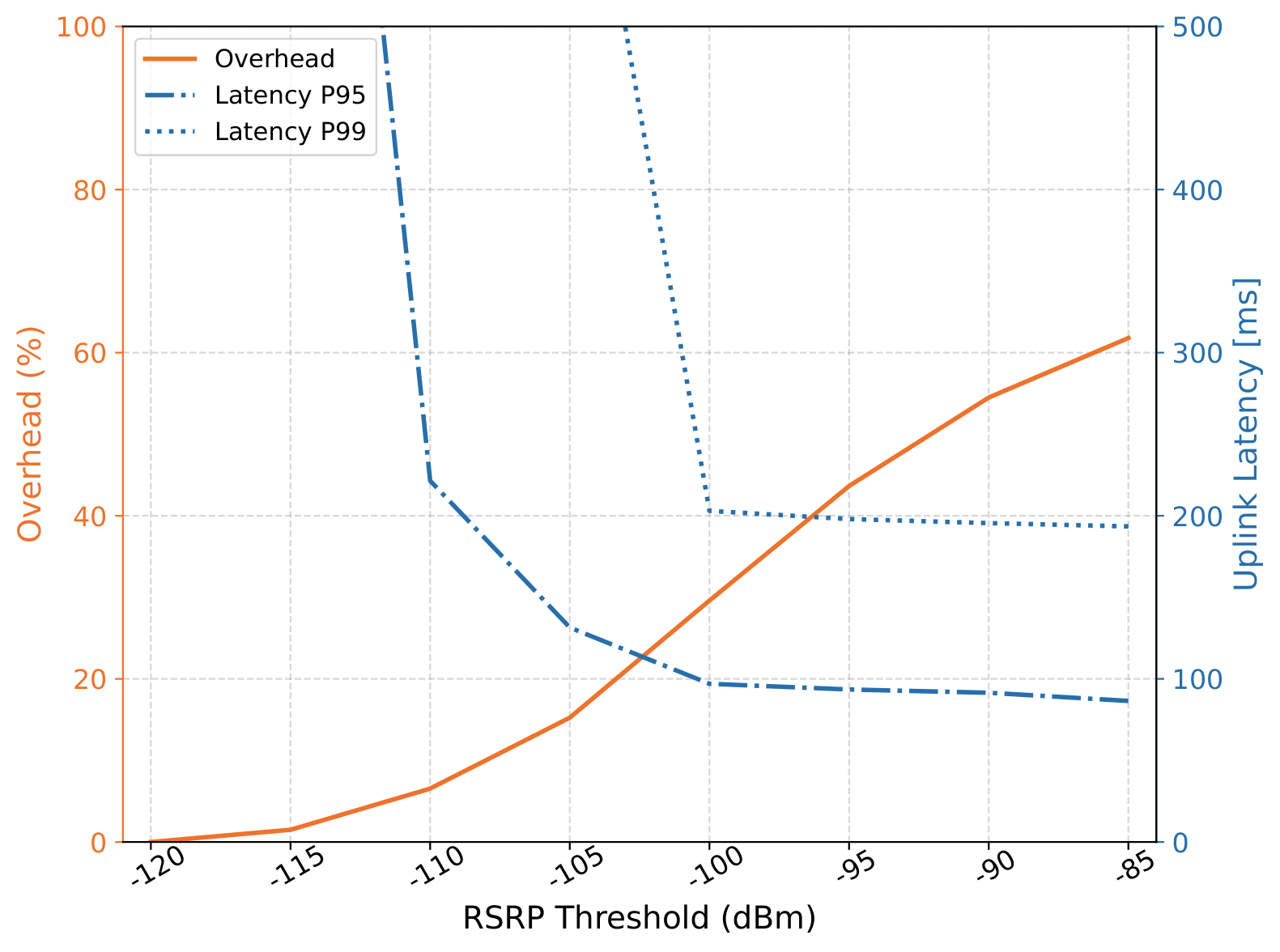}
    \caption{Trade-off between tail-latency and cost as a function of the RSRP threshold on UL latency percentiles.}
    \label{fig:sensitivity_rsrp}
\end{figure}

\begin{figure}[!h]
    \centering
    \includegraphics[width=0.95\columnwidth]{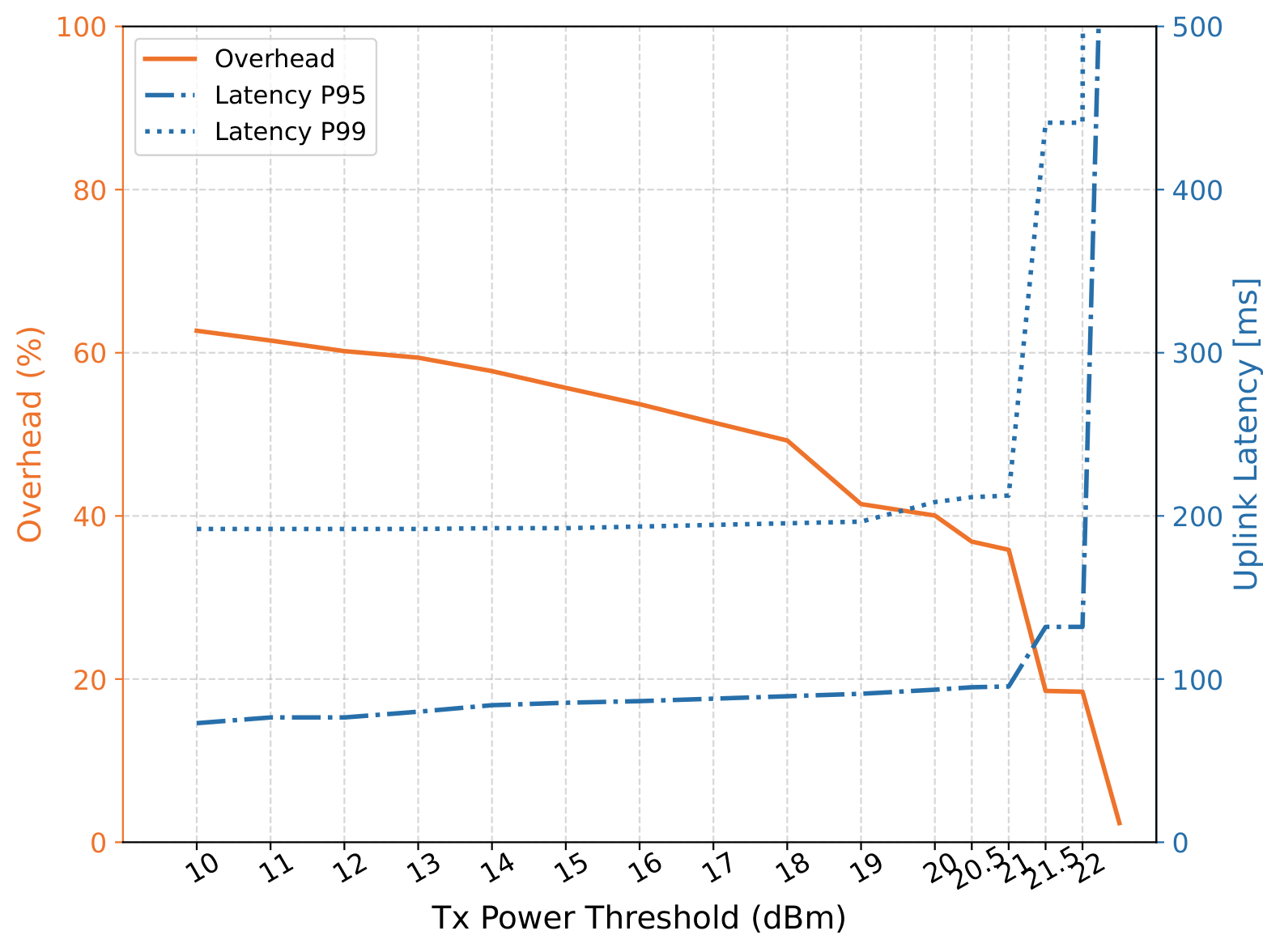}
    \caption{Trade-off between tail-latency and cost as a function of the UL Tx Pwr threshold on UL latency percentiles.}
    \label{fig:sensitivity_tx}
\end{figure}

Figs. \ref{fig:sensitivity_rsrp} and \ref{fig:sensitivity_tx} evaluate the performance of the PAAF algorithm by sweeping the RSRP and UL Tx Pwr thresholds, respectively, in order to identify operating points that avoid unnecessary duplication overhead while preserving latency-tail suppression. The results exhibit the expected monotonic behavior: relaxing the thresholds decreases the fraction of time the mechanism triggers, thereby reducing overhead, but it also allows more operation in severely degraded radio states, which re-inflates the 95th/99th percentile UL latency.

Following this analysis, the paper instantiates the PAAF controller with the thresholds balancing tail-latency containment against overhead in the rural trace $\theta_R=$-100~dBm, $\theta_U=$21~dBm, and $\theta_L=$150~ms.

\subsubsection{PAAF Switching:} 

Tables \ref{tab:paaf_switching_TELENOR_summary_4mbps} and \ref{tab:paaf_switching_TDC_summary_4mbps} summarize PAAF Switching at 4~Mbps, evaluated with each operator as the primary anchor. In all switching variants, only one interface carries the user-plane traffic at a time; consequently, the duplication overhead is considered 0\% as control channel overhead is negligible when compared to packet duplication, and the performance gains arise purely from exploiting inter-operator diversity.

\begin{table*}[!t]
\centering
\small
\setlength{\tabcolsep}{6pt}
\renewcommand{\arraystretch}{1.05}

\begin{adjustbox}{max width=\textwidth}
\begin{tabular}{lcccccc}
\hline\hline
\textbf{Mode} &
\textbf{MNO1 use [\%]} &
\textbf{MNO2 use [\%]} &
\textbf{Overhead [\%]} &
\textbf{UL Latency [ms] (90 \textbar\ 95 \textbar\ 99 \textbar\ 99.9\%)} &
\textbf{Late loss [\%]} ($>800$~ms) &
\textbf{True loss [\%]} ($=10^{4}$~ms) \\
\hline\hline

FD & 100 & 100 & 100 &
\ULlatPad{0.8em}{39}{53}{185}{481} &
$4\cdot 10^{-3}$ & $4\cdot 10^{-3}$ \\
\hline

RSRP - PAAF Switching & 90.4 & 9.6 & 0 & \ULlatPad{0.8em}{103}{281}{2497}{$10^4$} & 3.1 & 1.0 \\
UL Tx Pwr - PAAF Switching & 91.7 & 8.3 & 0 & \ULlatPad{0.8em}{96}{336}{$10^4$}{$10^4$} & 3.2 & 1.1 \\

\textbf{Latency - PAAF Switching} & \textbf{95.8} & \textbf{4.2} & 0 & \ULlatPad{0.8em}{\textbf{56}}{\textbf{152}}{\textbf{1099}}{\bm{$10^4$}} &   \textbf{1.4} & \textbf{0.5} \\
RSRP + UL Tx Pwr - PAAF Switching & 83.3 & 16.7 & 0 & \ULlatPad{0.8em}{122}{233}{1601}{$10^4$} & 2.2 & 0.7 \\

RSRP + Latency - PAAF Switching & 90.4 & 9.6 & 0 & \ULlatPad{0.8em}{103}{281}{2497}{$10^4$} & 3.1 & 1.0 \\

UL Tx Pwr + Latency - PAAF Switching & 89.3 & 10.7 & 0 & \ULlatPad{0.8em}{71}{223}{1588}{$10^4$} & 2.1 & 0.7 \\

RSRP + UL Tx Pwr + Latency - PAAF Switching & 83.3 & 16.7 & 0 & \ULlatPad{0.8em}{122}{233}{1601}{$10^4$} & 2.2 & 0.7 \\
\hline

Baseline (MNO1-only) & 100 & 0 & 0 & \ULlatPad{0.8em}{85}{283}{1952}{$10^4$} & 2.8 & 0.9 \\
\hline\hline
\end{tabular}
\end{adjustbox}

\caption{PAAF Switching performance overview of Experiment~2 (rural, 4~Mbps) with MNO1 as primary}
\label{tab:paaf_switching_TELENOR_summary_4mbps}
\end{table*}

\begin{table*}[!t]
\centering
\small
\setlength{\tabcolsep}{6pt}
\renewcommand{\arraystretch}{1.05}

\begin{adjustbox}{max width=\textwidth}
\begin{tabular}{lcccccc}
\hline\hline
\textbf{Mode} &
\textbf{MNO1 use [\%]} &
\textbf{MNO2 use [\%]} &
\textbf{Overhead [\%]} &
\textbf{UL Latency [ms] (90 \textbar\ 95 \textbar\ 99 \textbar\ 99.9\%)} &
\textbf{Late loss [\%]} ($>800$~ms) &
\textbf{True loss [\%]} ($=10^{4}$~ms) \\
\hline\hline

FD & 100 & 100 & 100 & \ULlatPad{0.8em}{39}{53}{185}{481} & $4\cdot 10^{-3}$ & $4\cdot 10^{-3}$ \\
\hline

RSRP - PAAF Switching & 14.0 & 86.0 & 0 & \ULlatPad{0.8em}{156}{924}{$10^4$}{$10^4$} & 5.2 & 2.5 \\

UL Tx Pwr - PAAF Switching & 16.5 & 83.5 & 0 & \ULlatPad{0.8em}{148}{683}{$10^4$}{$10^4$} & 4.7 & 2.5 \\

Latency - PAAF Switching & 8.8 & 91.2 & 0 & \ULlatPad{0.8em}{148}{953}{$10^4$}{$10^4$} & 5.4 & 2.3 \\


RSRP + UL Tx Pwr - PAAF Switching & 18.6 & 81.4 & 0 & \ULlatPad{0.8em}{161}{1015}{$10^4$}{$10^4$} & 5.7 & 2.4 \\

\textbf{RSRP + Latency - PAAF Switching} & \textbf{14.0} & \textbf{86.0} & \textbf{0} & \ULlatPad{0.8em}{\textbf{156}}{\textbf{924}}{\bm{$10^4$}}{\bm{$10^4$}} & \textbf{5.2} & \textbf{2.5} \\

UL Tx Pwr + Latency - PAAF Switching & 19.1 & 80.9 & 0 & \ULlatPad{0.8em}{199}{1378}{$10^4$}{$10^4$} & 7.1 & 3.1 \\

RSRP + UL Tx Pwr + Latency - PAAF Switching & 18.6 & 81.4 & 0 & \ULlatPad{0.8em}{161}{1015}{$10^4$}{$10^4$} & 5.8 & 2.4 \\
\hline

Baseline (MNO2-only) & 0 & 100 & 0 & \ULlatPad{0.8em}{457}{1941}{$10^4$}{$10^4$} & 8.8 & 4.6 \\
\hline\hline
\end{tabular}
\end{adjustbox}

\caption{PAAF Switching performance overview of Experiment~2 (rural, 4~Mbps) with MNO2 as primary}
\label{tab:paaf_switching_TDC_summary_4mbps}
\end{table*}

Both single-operator baselines exhibits a heavy latency tail and non-negligible losses where MNO2-only being substantially worse in this rural trace, confirming that the route contains extended segments where MNO2 becomes severely power-limited and unstable. FD provides an upper bound, drastically tightening the tail while virtually eliminating losses. Nonetheless, the overhead is 100\% compared with a single operator.

When compared to the single-operator baselines, Latency-based PAAF Switching as the only KPI consistently reduces the frequency and severity of extreme latency events while maintaining a latency tail substantially tighter than the baseline. When MNO1 is used as the primary anchor, latency-based switching improves the 95th-percentile UL latency and reduces late and true losses compared with the MNO1-only baseline. This improvement occurs because latency directly captures end-to-end service degradation, including queuing and retransmission effects that may not be fully reflected in slowly sampled radio KPIs. However, the 99th and 99.9th percentiles remain substantially higher than those achieved by full duplication, indicating that switching is not sufficient to suppress the most extreme tail events.


This limitation arises from the exclusive nature of switching. At each decision instant, only one interface carries the user traffic; therefore, the resulting latency distribution is governed by the selected link. If the controller switches to a weaker or already degraded alternative interface, the application is fully exposed to the tail behavior of that link. This is particularly relevant in the evaluated rural trace, where MNO2 exhibits stronger power-limited behavior and larger latency tails than MNO1. Consequently, radio-triggered switching may degrade performance when it redirects traffic from the more stable MNO1 link toward MNO2.

In contrast, radio-based switching policies based exclusively on, or combined with, RSRP and/or UL Tx Pwr degrade reliability when MNO1 is used as the primary anchor. As shown in Table \ref{tab:paaf_switching_TELENOR_summary_4mbps}, these policies increase both late-loss and true-loss ratios relative to the MNO1-only baseline and produce a substantially heavier high-percentile latency tail. Therefore, while switching can reduce duplication overhead to zero, it cannot exploit packet-level diversity in the same way as duplication. This explains why PAAF Switching can improve over some single-operator baselines but remains separated from PAAF PD and full duplication in terms of tail-latency suppression and loss reduction.


\begin{table*}[!t]
\centering
\small
\setlength{\tabcolsep}{6pt}
\renewcommand{\arraystretch}{1.05}

\begin{adjustbox}{max width=\textwidth}
\begin{tabular}{lcccccc}
\hline\hline
\textbf{Mode} &
\textbf{MNO1 use [\%]} &
\textbf{MNO2 use [\%]} &
\textbf{Overhead [\%]} &
\textbf{UL Latency [ms]} \textbf{(90 \textbar\ 95 \textbar\ 99\% \textbar\ 99.9\%)} &
\textbf{Late loss [\%]} ($>800$~ms) &
\textbf{True loss [\%]} ($=10^{4}$~ms) \\
\hline\hline
FD & 100 & 100 & 100  & 39 \textbar\ 53 \textbar\ 185 \textbar\ 481  & $4\cdot 10^{-3}$ & $4\cdot 10^{-3}$ \\

\hline

RSRP - PAAF PD                  & 100  & 27.2  & 27.2  & 44 \textbar\ 64 \textbar\ 197\textbar\ 489 & $2\cdot 10^{-2}$ & $2\cdot 10^{-2}$ \\

UL Tx Pwr - PAAF PD             & 100  & 33.5  & 33.5  & 43 \textbar\ 63 \textbar\ 213 \textbar\ 489  & $2\cdot 10^{-2}$ & $2\cdot 10^{-2}$ \\

\textbf{Latency - PAAF PD}               & \textbf{100}  & \textbf{8.4}  & \textbf{8.4}  & \textbf{48} \textbar\ \textbf{81} \textbar\ \textbf{207} \textbar\ \textbf{497} & \bm{$1\cdot 10^{-2}$} & \bm{$2\cdot 10^{-2}$} \\

RSRP AND UL Tx Pwr - PAAF PD        & 100  & 22.3  & 22.3  & 44 \textbar\ 66 \textbar\ 213 \textbar\ 489  & $2\cdot 10^{-2}$ & $2\cdot 10^{-2}$ \\


RSRP OR Latency - PAAF PD           & 100  & 27.6  & 27.6  & 44 \textbar\ 63 \textbar\ 191 \textbar\ 481  & $5\cdot 10^{-3}$ & $5\cdot 10^{-3}$ \\

UL Tx Pwr OR Latency - PAAF PD      & 100  & 34.1  & 34.1  & 43 \textbar\ 60 \textbar\ 191 \textbar\ 481  & $5\cdot 10^{-3}$ & $5\cdot 10^{-3}$ \\

\textbf{(RSRP AND UL Tx Pwr) OR Latency - PAAF PD}      & \textbf{100}  & \textbf{22.8}  & \textbf{22.8}  & \textbf{44} \textbar\ \textbf{63} \textbar\ \textbf{191} \textbar\ \textbf{481} & \bm{$5\cdot 10^{-3}$} & \bm{$5\cdot 10^{-3}$} \\


\hline
Baseline (MNO1-only)     & 100  & 0  & 0  & 85 \textbar\ 283 \textbar\ 1952 \textbar\ $10^{4}$ & 2.8 & 0.9 \\ 

\hline
\end{tabular}
\end{adjustbox}

\caption{PAAF PD performance overview of Experiment~2 (rural, 4~Mbps) with MNO1 as primary}
\label{tab:paaf_PD_TELENOR_summary_4mbps}
\end{table*}


\begin{table*}[!t]
\centering
\small
\setlength{\tabcolsep}{6pt}
\renewcommand{\arraystretch}{1.05}

\begin{adjustbox}{max width=\textwidth}
\begin{tabular}{lcccccc}
\hline\hline
\textbf{Mode} &
\textbf{MNO1 use [\%]} &
\textbf{MNO2 use [\%]} &
\textbf{Overhead [\%]} &
\textbf{UL Latency [ms]} \textbf{(90 \textbar\ 95 \textbar\ 99\% \textbar\ 99.9\%)} &
\textbf{Late loss [\%]} ($>800$~ms) &
\textbf{True loss [\%]} ($=10^{4}$~ms) \\
\hline\hline
FD & 100 & 100 & 100  & 39 \textbar\ 53 \textbar\ 185 \textbar\ 481   & $4\cdot 10^{-3}$ & $4\cdot 10^{-3}$ \\
\hline

RSRP - PAAF PD                  & 32.1  & 100  & 32.1  & 99 \textbar\ 130 \textbar\ 208 \textbar\ 484 & $2\cdot 10^{-2}$ & $2\cdot 10^{-2}$ \\

UL Tx Pwr - PAAF PD             & 38.2  & 100  & 38.2  & 87 \textbar\ 128 \textbar\ 211 \textbar\ 484 & $2\cdot 10^{-2}$ & $2\cdot 10^{-2}$ \\

\textbf{Latency - PAAF PD}               & \textbf{15.8}  & \textbf{100}  & \textbf{15.8}  & \textbf{115} \textbar\ \textbf{134} \textbar\ \textbf{231} \textbar\ \textbf{486} & \bm{$2\cdot 10^{-2}$} & \bm{$2\cdot 10^{-2}$} \\

RSRP AND UL Tx Pwr - PAAF PD        & 27.8  & 100  & 27.8  & 114 \textbar\ 133 \textbar\ 215 \textbar\ 484 & $2\cdot 10^{-2}$ & $2\cdot 10^{-2}$ \\


RSRP OR Latency - PAAF PD           & 34.0  & 100  & 34.0  & 65 \textbar\ 125 \textbar\ 203 \textbar\ 484 & $2\cdot 10^{-2}$ & $2\cdot 10^{-2}$ \\

UL Tx Pwr OR Latency - PAAF PD      & 40.2  & 100  & 40.2  & 60 \textbar\ 124 \textbar\ 202 \textbar\ 484 & $2\cdot 10^{-2}$ & $2\cdot 10^{-2}$ \\

\textbf{(RSRP AND UL Tx Pwr) OR Latency - PAAF PD}      & \textbf{29.8}  & \textbf{100}  & \textbf{29.8}  & \textbf{92} \textbar\ \textbf{127} \textbar\ \textbf{199} \textbar\ \textbf{484} & \bm{$1\cdot 10^{-2}$} & \bm{$1\cdot 10^{-2}$} \\


\hline
Baseline (MNO2-only)                    & 0  & 100  & 0  & 457 \textbar\ 1941  \textbar\ $10^{4}$ \textbar\ $10^{4}$ & 8.8 & 4.7 \\ 

\hline
\end{tabular}
\end{adjustbox}

\caption{PAAF PD performance overview of Experiment~2 (rural, 4~Mbps) with MNO2 as primary}
\label{tab:paaf_PD_TDC_summary_4mbps}
\end{table*}

\subsubsection{PAAF Partial Duplication (PD):} 

In contrast to PAAF Switching, PAAF PD preserves the primary link as the main transmission path and activates packet duplication over the secondary interface only when the primary link violates predefined performance thresholds. Consequently, PD aims to approximate the reliability of full duplication while substantially reducing the associated overhead. Tables \ref{tab:paaf_PD_TELENOR_summary_4mbps} and \ref{tab:paaf_PD_TDC_summary_4mbps} summarize the performance of PAAF PD when MNO1 and MNO2 are used as the primary anchor, respectively.

PAAF PD consistently performs close to FD while incurring only a fraction of the overhead required. Across all PD variants, the 95th- and 99th-percentile UL latency values are reduced by more than an order of magnitude compared to the baseline, and late-loss and true-loss ratios decrease from the percent range to the order of $10^{-2}$. 

Radio-triggered PD policies (RSRP- or UL Tx Pwr-based) activate duplication during approximately 22–28\% of the samples, yielding UL latency distributions that closely track those of FD.  

This behavior confirms that, unlike radio-only switching, radio KPIs are reliable indicators for duplication triggering when applied to a strong primary link, as they detect incipient link-budget stress without forcing traffic migration to a weaker alternative.

Latency-triggered PD activates less frequently (approximately 7–15\% overhead) and therefore incurs the lowest duplication cost. However, this reduction in overhead comes at the expense of a heavier extreme-latency tail, with the 99th- and 99.9th-percentile values, especially in MNO1 exceeding those achieved by radio-triggered PD. Logical combinations of radio and latency conditions provide intermediate operating points, allowing explicit tuning between reliability and overhead.

\subsection{Statistical Confidence of the Main Rural Replay Results}
\label{subsec:statistical_confidence}

To quantify the statistical stability of the main rural replay results, 99\% confidence intervals were computed with $\alpha=0.01$. For latency percentiles, the Dvoretzky--Kiefer--Wolfowitz (DKW) inequality was used to obtain a distribution-free confidence band for the ECDF. Given $n$ latency samples, the DKW confidence-band half-width is defined as
\begin{equation}
    \varepsilon_{\mathrm{DKW}}
    =
    \sqrt{\frac{\ln(2/\alpha)}{2n}} .
\end{equation}
Accordingly, the lower $F_{\mathrm{L}}(x)$ and upper $F_{\mathrm{U}}(x)$ DKW confidence bounds of the CDF are given by
\begin{equation}
    F_{\mathrm{L}}(x)
    =
    \max\left(\hat{F}_n(x)-\varepsilon_{\mathrm{DKW}},0\right),
\end{equation}
and
\begin{equation}
    F_{\mathrm{U}}(x)
    =
    \min\left(\hat{F}_n(x)+\varepsilon_{\mathrm{DKW}},1\right),
\end{equation}
where $\hat{F}_n(x)$ is the ECDF. Thus, with probability at least $1-\alpha$, the underlying CDF $F(x)$ lies within the interval $[F_{\mathrm{L}}(x),F_{\mathrm{U}}(x)]$ uniformly over all latency values. For the rural 4 Mbps replay trace, $N=311569$, which gives $\varepsilon_{\mathrm{DKW}}=3\cdot10^{-3}$. Hence, the empirical latency CDF is bounded within approximately $\pm 0.29$ percentage points at the 99\% confidence level. Since missing or non-received packets are encoded as the 10~s timeout value, the sample size $N$ corresponds to the number of transmitted packet opportunities rather than only successfully received packets. Consequently, this ensures that losses are incorporated into the empirical latency distribution instead of being removed from the percentile calculation.

Latency-domain confidence intervals for the reported percentiles were obtained by mapping the DKW probability band back to latency values through the empirical quantile function. For a target percentile probability $p$, the lower and upper percentile bounds are defined as
\begin{equation}
    q_{\mathrm{L}}(p)
    =
    \hat{Q}\left(\max(p-\varepsilon_{\mathrm{DKW}},0)\right),
\end{equation}
and
\begin{equation}
    q_{\mathrm{U}}(p)
    =
    \hat{Q}\left(\min(p+\varepsilon_{\mathrm{DKW}},1)\right),
\end{equation}
where $\hat{Q}(\cdot)$ denotes the empirical quantile function. Therefore, the latency-domain confidence interval for the percentile $p$ is $[q_{\mathrm{L}}(p),q_{\mathrm{U}}(p)]$. For packet-level proportions, including packets below 150 ms, late loss, true loss, interface usage, and duplication overhead, Wilson score intervals were used. This avoids Gaussian assumptions and is appropriate for the heavy-tailed latency distributions observed in the rural traces.

Table~\ref{tab:ci_rural_4mbps} reports the resulting confidence intervals for the main rural 4 Mbps replay comparison. The results show that the improvement of PAAF PD over the single-operator baselines is statistically stable. In particular, the $P_{95}$ and $P_{99}$ latency intervals of PAAF PD remain well below those of MNO1-only and MNO2-only, while remaining close to FD. At the same time, PAAF PD achieves this performance with substantially lower duplication overhead than FD.

\begin{table*}[t]
\centering
\renewcommand{\arraystretch}{1.15}
\scriptsize
\begin{adjustbox}{max width=\textwidth}
\begin{tabular}{lcccc}
\hline
\textbf{Strategy} &
\textbf{$P_{95}$ UL [ms]} &
\textbf{$P_{99}$ UL [ms]} &
\textbf{$\leq 150$ ms [\%]} &
\textbf{Late loss [\%]} \\
\hline
MNO1-only &
283 (251--321) &
1952 (1514--$10^{4}$) &
91.6 (91.5--91.7) &
2.8 (2.7--2.9) \\

MNO2-only &
1941 (1875--2117) &
$10^{4}$ ($10^{4}$--$10^{4}$) &
84.7 (84.5--84.9) &
8.7 (8.6--8.9) \\

PAAF Switching, MNO1 anchor &
192 (180--207) &
1105 (911--1475) &
93.8 (93.7--93.9) &
1.4 (1.3--1.5) \\

PAAF Switching, MNO2 anchor &
1265 (1162--1413) &
$10^{4}$ ($10^{4}$--$10^{4}$) &
89.1 (88.9--89.2) &
6.7 (6.6--6.8) \\

PAAF PD, MNO1 anchor &
63 (59--67) &
191 (171--227) &
98.3 (98.2--98.3) &
$5\cdot10^{-3}$ ($3\cdot10^{-3}$--$1\cdot10^{-2}$) \\

PAAF PD, MNO2 anchor &
126 (125--128) &
197 (180--237) &
97.9 (97.8--97.9) &
$2\cdot10^{-2}$ ($1\cdot10^{-2}$--$2\cdot10^{-2}$) \\

FD &
53 (51--57) &
185 (166--220) &
98.4 (98.3--98.4) &
$4\cdot10^{-3}$ ($2\cdot10^{-3}$--$8\cdot10^{-3}$) \\
\hline
\end{tabular}
\end{adjustbox}
\caption{Confidence intervals for the main rural 4 Mbps replay comparison. Values in parentheses indicate 99\% confidence intervals. PAAF Switching is according to Latency and PAAF PD is according to (RSRP AND UL Tx Pwr) OR Latency metrics}
\label{tab:ci_rural_4mbps}
\end{table*}

\section{Discussion} \label{Sec:Discussion}

This section discusses the implications of the experimental findings, with particular emphasis on the limitations of conventional multi-connectivity strategies, the suitability of the proposed PAAF mechanisms for rural and industrial deployments, and the efficiency–reliability trade-offs revealed by the normalized overhead analysis.

\subsection{Limitations of Link Aggregation in Co-Located Rural Deployments}

One of the initial hypotheses of this work was that link aggregation could offer a favorable compromise between reliability and efficiency by splitting traffic across multiple interfaces without incurring duplication overhead. However, the experimental results demonstrate that, in the evaluated rural scenario, link aggregation fails to meet the stringent latency and reliability requirements of industrial teleoperation.

This outcome shows that one of the main limiting factors to link aggregation is related to UL Tx Pwr-control constraints. In principle, reducing the target data rate per interface should relax the required SINR, thereby improving robustness. However, the measurements show that, in the rural scenario, the UE frequently operates in a power-limited regime, with UL Tx Pwr saturating near $P_{MAX, UE}$. 
Moderate reductions in target data rate do not translate into proportional improvements in SINR margin, as the scheduler adapts resource allocation rather than significantly lowering the required power spectral density. Consequently, traffic splitting does not eliminate the underlying bottleneck, and aggregation fails to suppress extreme latency events in coverage-limited segments.

\subsection{Operator Asymmetry and Correlated Fading Interpretation}
The effectiveness of selection-based switching is constrained by both operator asymmetry and by the degree of correlation between the two access paths. At the Aalborg municipal level, approximately 45\% of antenna sites are shared between MNO1 and MNO2, according to \cite{mastedatabasen2026}. However, along the evaluated rural trajectory, MNO1 generally provides the more stable uplink, whereas MNO2 exhibits stronger power-limited behavior, larger latency tails, and higher late-loss and true-loss ratios. All MNO2 gNBs encountered during the measurement campaign were deployed at sites physically co-located with MNO1 infrastructure, as illustrated in Fig.~\ref{fig:rural_map}. This deployment condition substantially increases the likelihood of correlated large-scale fading and shared coverage holes along the route. Under such conditions, the probability that both links simultaneously experience degradation during the same decision window remains non-negligible. Consequently, switching between interfaces may not provide sufficient diversity gain when impairments are spatially correlated.

This behavior also explains why radio-based switching and radio-triggered partial duplication should be interpreted differently. Radio KPIs such as RSRP and UL Tx Pwr can indicate coverage degradation or power-limited operation, but in a switching policy they are used for exclusive path selection. Therefore, when MNO1 is the primary anchor, radio-triggered switching may redirect traffic toward MNO2 during periods in which the secondary link is less reliable, exposing the application entirely to the tail behavior of the selected interface. Thus, the limited performance of radio-triggered switching in this trace should not be interpreted as evidence that radio KPIs are generally uninformative, but rather as a consequence of using them for exclusive interface selection under asymmetric and partially correlated link conditions.

In contrast, duplication reduces the effective outage probability to the probability that both links are simultaneously degraded at the packet level. Even under correlated fading, short-term channel fluctuations, scheduler behavior, and independent retransmission processes introduce residual diversity. Conditional duplication therefore remains effective in suppressing the upper tail of the latency distribution, even when spatial diversity is limited.

This behavior is consistent with the operator-asymmetry interpretation discussed above: switching is particularly sensitive to the quality of the alternative link because it abandons the primary path rather than combining both paths.

\subsection{Practical Implications for Remote Monitoring, Teleoperation and Emergency Response}

The results have direct implications for mission-critical applications such as industrial automation, remote monitoring, teleoperation, public safety, and emergency response, which share stringent latency constraints, intolerance to bursty delay excursions, and sensitivity to packet loss. 

The experiments reveal that UL impairments dominate performance degradation in rural scenarios, driven by power-limited transmission, handovers, and abrupt radio-condition transitions. Even when median latency remains low, rare but severe latency spikes can violate service requirements and destabilize control applications. Therefore, solutions that focus solely on average performance or throughput are insufficient; instead, mechanisms must explicitly suppress extreme tail behavior.

From this perspective, redundancy-aware multi-connectivity remains essential for industrial-grade reliability in rural deployments, but it must be applied selectively and intelligently to avoid excessive resource consumption.

\subsection{Control-Loop Interpretation of Switching vs PD}
 
Switching operates as a periodically evaluated selection-diversity mechanism, with reaction time bounded by the KPI observation delay plus the heartbeat interval $T_s$. As a result, transient impairments occurring within a decision window cannot be mitigated until the next evaluation instant. In rural deployments, retransmission bursts and transient power-limited episodes often evolve on time scales comparable to or shorter than $T_s$, such that short-lived latency spikes may remain undetected. Conversely, longer-lasting degradations can be mitigated only after a bounded reaction delay, during which extreme latency excursions may already occur. This discrete-time quantization therefore limits the effectiveness of switching in suppressing both fast transient spikes and the onset of sustained outages.

In contrast, PD is event-driven with respect to degradation detection and does not incur additional decision quantization delay. Although its activation remains constrained by KPI observation latency, redundancy can be enabled immediately once degradation is detected. This asymmetry in control timing partially explains the superior tail-latency suppression observed for PD relative to switching.

\subsection{Normalized Overhead and Efficiency Considerations}

To assess the efficiency implications of redundancy mechanisms, a normalized overhead model was introduced in which traffic transmitted over the secondary interface is weighted by a multiplicative factor. This abstraction captures relative resource expenditure, such as network load, without relying on operator-specific pricing assumptions, and enables a technology-agnostic comparison of redundancy strategies.

\begin{figure}[h]
    \centering
    \includegraphics[width=\columnwidth]{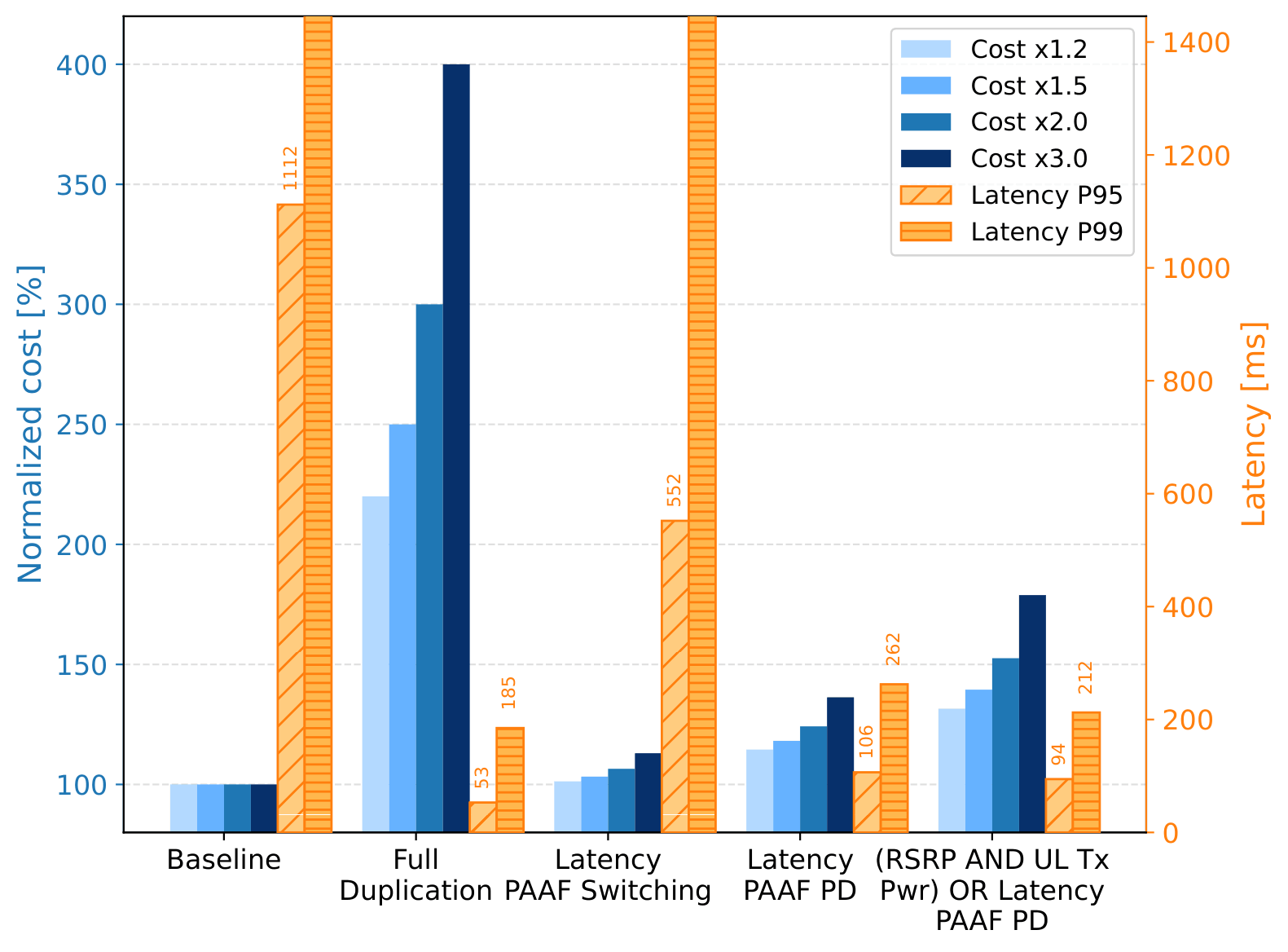}
    \caption{Summary of the latency–overhead trade-off for the evaluated MC strategies in the rural 4 Mbps scenario}
    \label{fig:cost_comparison}
\end{figure}

Fig.~\ref{fig:cost_comparison} illustrates the trade-off between UL latency performance and normalized overhead for the evaluated strategies, considering multiple weighting factors for the secondary link ($\times$1.2, $\times$1.5, $\times$2, and $\times$3). The latency metric reflects the 95th- and 99th-percentile UL latency aggregated across both operators. The resulting comparison reveals a clear hierarchy among the strategies. FD achieves the strongest reliability, with consistently low high-percentile latency, but incurs a linear increase in overhead, effectively doubling the transmitted traffic independently of channel conditions. At the opposite extreme, PAAF Switching introduces negligible overhead, yet provides only limited protection against extreme latency events, particularly in scenarios with correlated link degradation. PAAF PD occupies an intermediate and highly favorable operating region, achieving latency performance close to FD while reducing the normalized overhead by up to approximately 75\% relative to unconditional duplication.

Fig. \ref{fig:cost_comparison} summarizes the latency–overhead trade-off already quantified in Tables \ref{tab:paaf_switching_TELENOR_summary_4mbps}-\ref{tab:paaf_PD_TDC_summary_4mbps}. FD provides the strongest tail-latency suppression but incurs the maximum redundancy overhead. PAAF Switching introduces negligible duplication overhead, but its ability to suppress extreme latency events remains limited in the evaluated rural scenario. PAAF PD occupies an intermediate operating region, retaining most of the reliability gain of FD while substantially reducing redundancy overhead.

\subsection{PAAF Partial Duplication as a Practical Redundancy Mechanism}

Among all evaluated approaches within the evaluated rural scenario, within the evaluated rural 4 Mbps scenario, PAAF PD provides the most favorable reliability–overhead trade-off among the tested strategies. By preserving a cost-preferred primary link and activating duplication only during periods of demonstrable link stress, PD absorbs abrupt retransmission bursts and transient outages that switching alone cannot mitigate.

A key insight of this work is that radio KPIs such as RSRP and UL Tx Pwr are particularly effective triggers for PD, even though they are insufficient, or even detrimental, for switching decisions. When used for PD, these KPIs provide early warning of imminent power-limited operation, enabling duplication to be activated during early phases of link degradation, often before severe latency spikes fully develop. This reduces reliance on latency-only feedback, lowers system uncertainty, and shortens the time-to-trigger compared to purely reactive schemes.

In coverage-constrained deployments with partial correlation, such as the rural scenario, PD therefore retains a meaningful reduction in tail probability, even when switching-based strategies provide limited benefit. In contrast, switching selects a single interface at each decision instant, and its outage probability remains governed by the tail probability of the selected link, without benefiting from multiplicative suppression of simultaneous impairments. Moreover, switching decisions are subject to periodic evaluation intervals, which further limit their ability to suppress short-lived but severe latency excursions. 

The empirical results observed along the rural trajectory align with this structural diversity argument, as duplication suppresses joint exceedance events while switching remains limited to single-link performance. From a system-level perspective, PD implicitly balances tail-latency reduction against duplication overhead, approximating a reliability–resource trade-off without requiring centralized optimization.

\begin{figure}[h]
\centering
\includegraphics[width=\columnwidth]{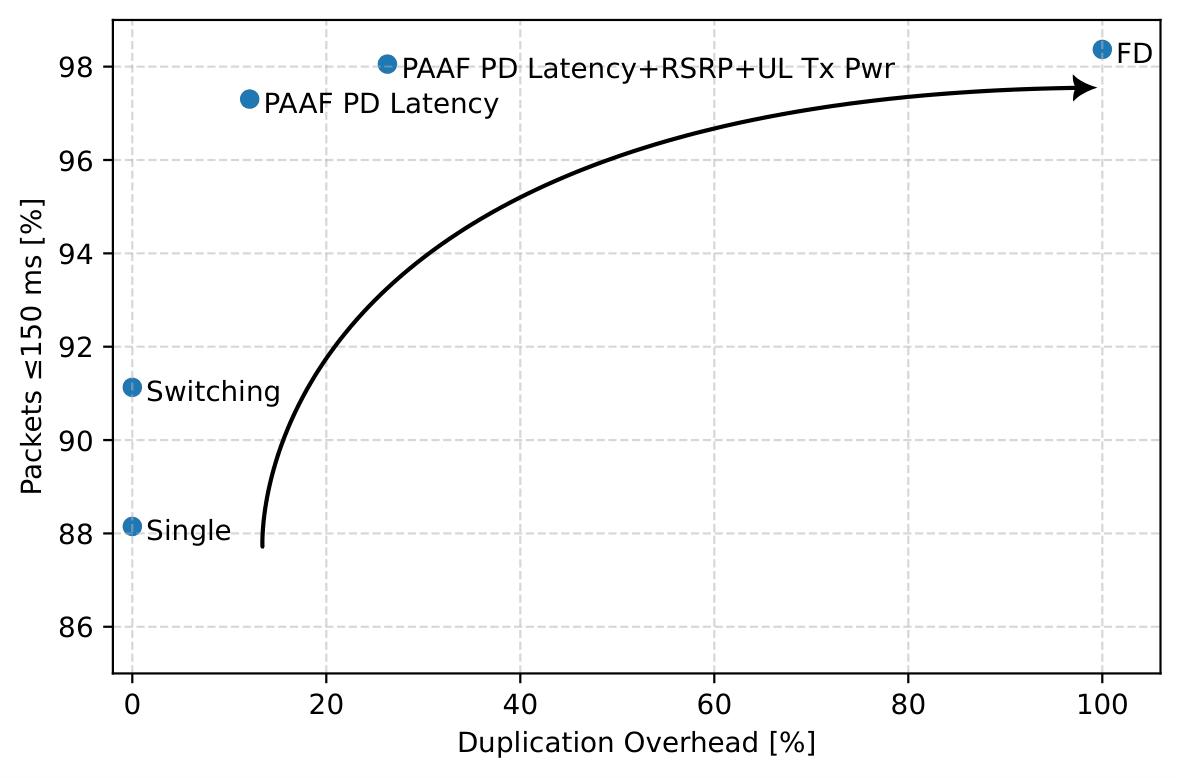}
\caption{Reliability–overhead trade-off under rural 4 Mbps operation, expressed as the fraction of packets satisfying the 150 ms uplink-latency target versus the duplication overhead}
\label{fig:pareto_tradeoff}
\end{figure}

Fig.~\ref{fig:pareto_tradeoff} summarizes the reliability–overhead positioning of the evaluated strategies under the 150~ms constraint. 

A clear trade-off emerges: increasing redundancy improves reliability but incurs additional overhead. PAAF switching achieves a substantial reliability improvement without duplication overhead, occupying the low-cost region, whereas PAAF PD with different strategies achieves additional tail suppression at moderate cost. Notably, PAAF PD appears near the knee of the trade-off curve, beyond which further reliability gains require disproportionately higher overhead. FD yields only marginal improvement beyond PD despite incurring maximum overhead.

\subsection{Limitations of the Study}

While the presented results provide extensive empirical insight into multi-connectivity behavior under commercial 5G deployments, several limitations must be acknowledged.

First, the measurement campaign was conducted in Denmark using two commercial mobile network operators under typical 5G NR spectrum allocations (including sub-6 GHz bands) and a 5G NSA architecture. Consequently, the findings reflect the deployment characteristics, network configurations, and spectrum usage of these operators and may not generalize directly to SA 5G deployments, mmWave systems, or networks operating under different spectrum and densification regimes.

Second, the rural scenario analyzed in detail exhibits partial infrastructure co-location between operators, which results in correlated large-scale fading and coverage holes. While this configuration is representative of many sparsely populated regions where tower sharing is common, the degree of spatial correlation may differ in areas with greater inter-operator site diversity. Therefore, the relative effectiveness of switching versus duplication strategies may vary depending on the spatial independence of the access paths.

Finally, the experiments focus on constant-bit-rate UDP telemetry traffic to enable controlled latency characterization. Although this traffic model is representative of real-time industrial telemetry and teleoperation streams, adaptive bitrate applications or bursty traffic patterns may exhibit different dynamics, particularly under scheduler-driven resource allocation. Therefore, the reported gains should be interpreted as representative of the evaluated dual-operator rural 5G NSA deployment rather than as universal performance bounds for all 5G multi-connectivity configurations.

Despite these limitations, the measurement campaign captures realistic mobility, commercial scheduling behavior, and uplink power-control dynamics, providing practical insight into multi-connectivity trade-offs in coverage-limited environments.

\section{Conclusion and Future work} \label{Sec:Conclusion_Future_Work}

This paper presented a comprehensive experimental investigation of multi-connectivity strategies for reliable low-latency communications over commercial 5G Non-Standalone networks. Based on extensive measurement campaigns conducted across urban, suburban, and rural scenarios, the study analyzed the joint behavior of radio KPIs, uplink power control, latency distributions, and packet loss, and evaluated the effectiveness of duplication, aggregation, and adaptive failover mechanisms under realistic operating conditions.

The multi-scenario results of Experiment~1 revealed that latency and reliability in commercial 5G networks cannot be inferred solely from coverage indicators such as RSRP, and that their root causes vary significantly across environments. While urban and suburban scenarios exposed the influence of operator-specific configurations and link-budget limitations, the rural scenario emerged as the most critical case, characterized by power-limited UL operation, correlated impairments due to co-located base stations, and severe latency outliers. In particular, the rural measurements showed that moderate data-rate reductions did not yield the expected latency improvements, as the UE frequently operated near its maximum transmit power, limiting the effectiveness of theoretical SINR gains. These observations directly motivated Experiment 2, which focused on the rural scenario to investigate whether UL data rate adaptation and UL Tx Pwr provide actionable indicators for adaptive switching and duplication strategies. The results suggest that UL Tx Pwr and latency dynamics offer more reliable guidance than data rate alone for triggering multi-connectivity control mechanisms under coverage-limited conditions.

Nevertheless, Experiment~2 demonstrates that at sufficiently low data rates (0.5 and 0.25~Mbps), the rural route becomes largely compatible with the latency requirements of remote video monitoring and teleoperation. In contrast, at higher rates such as 4~Mbps, a non-negligible fraction of time exhibits latency violations for at least one operator, and occasionally for both. These results indicate that sustaining high UL data rates in rural coverage conditions requires either conservative bitrate adaptation or the exploitation of redundancy through multi-connectivity.

Within this context, the proposed PAAF framework addresses the shortcomings of static multi-connectivity approaches. The experimental evaluation shows that link aggregation was ineffective in the evaluated rural deployment due to power-limited UL operation and reduced spatial diversity, while switching-based strategies, although beneficial compared to single-operator baselines, remain constrained by correlated link degradation and limited reaction speed. Under the evaluated rural 5G NSA conditions, PAAF PD provides strong reliability improvements relative to single-operator baselines and PAAF Switching. By activating redundancy selectively, PAAF PD achieves latency and packet-loss performance close to full duplication while reducing the associated overhead by up to approximately 75\%.

A key insight of this work is that radio KPIs such as RSRP and UL Tx Pwr, while insufficient or even detrimental when used alone for switching decisions, are highly effective triggers for PD. When combined with latency feedback, these indicators enable proactive redundancy activation, reduce reliance on purely reactive control channels, shorten the time-to-trigger, and improve robustness against abrupt radio degradations, handover failures, and technology fallback. This suggests that PAAF PD may be suitable for safety-critical and mission-critical IIoT applications, where resilience to rare but disruptive events is as important as average performance.

Future work will extend the PAAF framework toward fully adaptive multi-connectivity controllers that jointly optimize PD and load balancing. Promising directions include dynamic traffic splitting across operators based on fused radio and latency KPIs, smooth transitions between balanced and redundant transmission modes, and tighter integration with application-layer bitrate adaptation. Evaluating such mechanisms under heterogeneous access technologies and at larger scales constitutes an important step toward resilient, efficient, and deployable connectivity solutions for 5G and beyond networks.

\section*{Acknowledgment}

The authors gratefully acknowledge the support of Sebastian Bro Damsgaard, Carmen Covadonga Blanco Gonzalez and Per Hartmann Christensen during the measurement campaign and the setup indications.

\bibliographystyle{IEEEtran}
\bibliography{References}

\end{document}